\documentclass[12pt]{article}        
\newcommand{\dd}{\displaystyle}
\newcommand{\nn}{\nonumber}
\newcommand{\smallz}{{\scriptscriptstyle Z}} 
\newcommand{\smallw}{{\scriptscriptstyle W}} %
\newcommand{\smallh}{{\scriptscriptstyle H}} %
\newcommand{\eps}{\epsilon}
\newcommand{\sic}{ {\hat s} }
\newcommand{\coc}{ {\hat c} }
\newcommand{\acur}{ {\hat \alpha} }
\newcommand{\mz}{M_\smallz}
\newcommand{\mw}{M_\smallw}
\newcommand{\mh}{M_\smallh}
\newcommand{\mt}{m_t}
\newcommand{\ft}{\footnotesize}
\newcommand{\sineff}{\mbox{$\sin^2 \theta^{{ lept}}_{{ eff}}$} }
\newcommand{\scur}{\mbox{$\hat{s}^2$}}
\newcommand{\sincur}{\mbox{$\sin^{2}\!\hat{\theta}_{\scriptscriptstyle W} 
                           (\mz^2)$}}
\newcommand{\ccur}{\mbox{$\hat{c}^2$}}
\newcommand{\drho}{\mbox{$ \Delta \rho$}}
\newcommand{\drhoc}{\mbox{$ \Delta \hat{\rho}$}}
\newcommand{\gmu}{\mbox{$ G_\mu $}}
\newcommand{\gmtd}{\mbox{$ O(G_\mu^2 m_t^2 \mz^2) $}}
\newcommand{\amtd}{\mbox{$ O(\alpha^2 m_t^2 /\mw^2) $}}
\newcommand{\amtq}{\mbox{$ O(\alpha^2 m_t^4 /\mw^4) $}}
\newcommand{\gmtq}{\mbox{$ O(G_\mu^2 m_t^4) $}}
\newcommand{\ew}{electroweak}
\newcommand{\msbar}{\overline{\rm MS}}
\def\lequiv{\raise 0.4ex \hbox{$<$} \kern -0.8 em 
\lower 0.62 ex \hbox{$\sim$}}
\def\gequiv{\raise 0.4ex \hbox{$>$} \kern -0.7 em
 \lower 0.62 ex \hbox{$\sim$}}
\newcommand{\equ}[1]{Eq.~(\ref{#1})}
\newcommand{\eqs}[1]{Eqs.~(\ref{#1})}
\newcommand{\be}{\begin{equation}}
\newcommand{\ee}{\end{equation}}
\newcommand{\bea}{\begin{eqnarray}}
\newcommand{\eea}{\end{eqnarray}}
\renewcommand{\thefootnote}{\fnsymbol{footnote} }
\begin{document}

\begin{flushright}
        \small
        MPI-PhT-96-85\\
        August 1996
\end{flushright}
\begin{center}
{\large\bf 
Two-loop heavy top effects   \\
           on precision observables\footnote{Invited talk at
 the Third International Symposium on 
Radiative Corrections,
Cracow, Poland, 1-5 August 1996. 
}}
 \\
\vspace{.62cm}
{\sc\small  Paolo Gambino}\\
\vspace{.2cm}
{\em\small Max Planck Institut f\"ur Physik, Werner Heisenberg Institut,\\
 F\"ohringer Ring 6, D80805 M\"unchen, Germany}
\end{center}
\vspace{.2cm} 
\begin{center}
{\bf\small Abstract}
\end{center}
\vspace{-.7cm}
\begin{quotation}
\small
\noindent
The corrections induced by a heavy top on the main precision observables 
are now available up to $O(\alpha^2 \mt^2/\mw^2)$.
The new results  
 imply a significant reduction of the 
theoretical uncertainty and can have a sizable
impact on the determination of \sineff.
\end{quotation}
\vspace{.5cm}
\noindent
The very precise measurements carried out at LEP and SLC in the recent past
have made the study of higher order radiative corrections 
necessary in order to test the Standard Model, 
and possibly to  uncover hints of new physics.
 The one-loop corrections to all the relevant
\ew\ observables are by now 
very well established \cite{YB},
and  two and three-loop effects have been investigated in several  cases. 
The dominance of a heavy top quark in the one-loop \ew\ corrections, 
which depend quadratically on the top mass, has allowed to predict
with good approximation
the  mass of the heaviest quark before its actual discovery. 

Among the 
higher order effects connected with these large non-decoupling
contributions, the QCD corrections are now known 
through $O(\alpha_s^2)$ \cite{QCD}.
As for the  purely \ew\
effects originated at higher orders by the large Yukawa coupling 
of the top,
reducible contributions  have been first studied in 
\cite{CHJ}, while a  thorough investigation of leading irreducible 
two-loop
contributions has been initiated in \cite{van},
 in the limit of a
massless Higgs, and later continued by Barbieri {\em et al.} and others
for arbitrary $\mh$ \cite{barb}.
In this talk I will illustrate 
some implications of the new calculation of the \amtd\
corrections to the main precision observables \cite{physlett,zako}.

The result of the calculation of the leading \gmtq\ effects 
on the $\rho$
parameter \cite{barb}  is shown in Fig.1 (upper curve). The 
correction  is relatively sizable
and in the  heavy Higgs case 
 reaches the permille level in the prediction of $\sineff$, 
comparable to the present experimental accuracy \cite{war}.
 \begin{figure}
\setlength{\unitlength}{0.240900pt}
\ifx\plotpoint\undefined\newsavebox{\plotpoint}\fi
\sbox{\plotpoint}{\rule[-0.175pt]{0.350pt}{0.350pt}}%
\begin{picture}(1200,800)(0,0)
\sbox{\plotpoint}{\rule[-0.175pt]{0.350pt}{0.350pt}}%
\put(264,158){\rule[-0.175pt]{0.350pt}{151.526pt}}
\put(264,215){\rule[-0.175pt]{4.818pt}{0.350pt}}
\put(242,215){\makebox(0,0)[r]{\ft -5}}
\put(1416,215){\rule[-0.175pt]{4.818pt}{0.350pt}}
\put(264,330){\rule[-0.175pt]{4.818pt}{0.350pt}}
\put(242,330){\makebox(0,0)[r]{\ft -4}}
\put(1416,330){\rule[-0.175pt]{4.818pt}{0.350pt}}
\put(264,444){\rule[-0.175pt]{4.818pt}{0.350pt}}
\put(242,444){\makebox(0,0)[r]{\ft -3}}
\put(1416,444){\rule[-0.175pt]{4.818pt}{0.350pt}}
\put(264,558){\rule[-0.175pt]{4.818pt}{0.350pt}}
\put(242,558){\makebox(0,0)[r]{\ft -2}}
\put(1416,558){\rule[-0.175pt]{4.818pt}{0.350pt}}
\put(264,673){\rule[-0.175pt]{4.818pt}{0.350pt}}
\put(242,673){\makebox(0,0)[r]{\ft -1}}
\put(1416,673){\rule[-0.175pt]{4.818pt}{0.350pt}}
\put(264,787){\rule[-0.175pt]{4.818pt}{0.350pt}}
\put(242,787){\makebox(0,0)[r]{\ft 0}}
\put(1416,787){\rule[-0.175pt]{4.818pt}{0.350pt}}
\put(264,158){\rule[-0.175pt]{0.350pt}{4.818pt}}
\put(264,113){\makebox(0,0){\ft 0}}
\put(264,767){\rule[-0.175pt]{0.350pt}{4.818pt}}
\put(431,158){\rule[-0.175pt]{0.350pt}{4.818pt}}
\put(431,113){\makebox(0,0){\ft 100}}
\put(431,767){\rule[-0.175pt]{0.350pt}{4.818pt}}
\put(599,158){\rule[-0.175pt]{0.350pt}{4.818pt}}
\put(599,113){\makebox(0,0){\ft 200}}
\put(599,767){\rule[-0.175pt]{0.350pt}{4.818pt}}
\put(766,158){\rule[-0.175pt]{0.350pt}{4.818pt}}
\put(766,113){\makebox(0,0){\ft 300}}
\put(766,767){\rule[-0.175pt]{0.350pt}{4.818pt}}
\put(934,158){\rule[-0.175pt]{0.350pt}{4.818pt}}
\put(934,113){\makebox(0,0){\ft 400}}
\put(934,767){\rule[-0.175pt]{0.350pt}{4.818pt}}
\put(1101,158){\rule[-0.175pt]{0.350pt}{4.818pt}}
\put(1101,113){\makebox(0,0){\ft 500}}
\put(1101,767){\rule[-0.175pt]{0.350pt}{4.818pt}}
\put(1269,158){\rule[-0.175pt]{0.350pt}{4.818pt}}
\put(1269,113){\makebox(0,0){\ft 600}}
\put(1269,767){\rule[-0.175pt]{0.350pt}{4.818pt}}
\put(1436,158){\rule[-0.175pt]{0.350pt}{4.818pt}}
\put(1436,113){\makebox(0,0){\ft 700}}
\put(1436,767){\rule[-0.175pt]{0.350pt}{4.818pt}}
\put(264,158){\rule[-0.175pt]{282.335pt}{0.350pt}}
\put(1436,158){\rule[-0.175pt]{0.350pt}{151.526pt}}
\put(264,787){\rule[-0.175pt]{282.335pt}{0.350pt}}
\put(45,472){\makebox(0,0)[l]{\shortstack{$\Delta\hat{\rho}^{(2,top)}$}}}
\put(850,68){\makebox(0,0){\small $\mh ({\rm GeV})$}}
\put(264,158){\rule[-0.175pt]{0.350pt}{151.526pt}}
\sbox{\plotpoint}{\rule[-0.500pt]{1.000pt}{1.000pt}}%
\put(1306,722){\makebox(0,0)[r]{{\ft heavy higgs exp}}}
\put(1328,722){\rule[-0.500pt]{15.899pt}{1.000pt}}
\put(425,334){\usebox{\plotpoint}}
\put(425,334){\rule[-0.500pt]{1.032pt}{1.000pt}}
\put(429,333){\rule[-0.500pt]{1.032pt}{1.000pt}}
\put(433,332){\rule[-0.500pt]{1.032pt}{1.000pt}}
\put(437,331){\rule[-0.500pt]{1.032pt}{1.000pt}}
\put(442,330){\rule[-0.500pt]{1.032pt}{1.000pt}}
\put(446,329){\rule[-0.500pt]{1.032pt}{1.000pt}}
\put(450,328){\rule[-0.500pt]{1.032pt}{1.000pt}}
\put(454,327){\usebox{\plotpoint}}
\put(458,326){\usebox{\plotpoint}}
\put(462,325){\usebox{\plotpoint}}
\put(465,324){\usebox{\plotpoint}}
\put(469,323){\usebox{\plotpoint}}
\put(473,322){\usebox{\plotpoint}}
\put(476,321){\usebox{\plotpoint}}
\put(480,320){\usebox{\plotpoint}}
\put(484,319){\rule[-0.500pt]{1.032pt}{1.000pt}}
\put(488,318){\rule[-0.500pt]{1.032pt}{1.000pt}}
\put(492,317){\rule[-0.500pt]{1.032pt}{1.000pt}}
\put(496,316){\rule[-0.500pt]{1.032pt}{1.000pt}}
\put(501,315){\rule[-0.500pt]{1.032pt}{1.000pt}}
\put(505,314){\rule[-0.500pt]{1.032pt}{1.000pt}}
\put(509,313){\rule[-0.500pt]{1.032pt}{1.000pt}}
\put(513,312){\rule[-0.500pt]{1.032pt}{1.000pt}}
\put(518,311){\rule[-0.500pt]{1.032pt}{1.000pt}}
\put(522,310){\rule[-0.500pt]{1.032pt}{1.000pt}}
\put(526,309){\rule[-0.500pt]{1.032pt}{1.000pt}}
\put(531,308){\rule[-0.500pt]{1.032pt}{1.000pt}}
\put(535,307){\rule[-0.500pt]{1.032pt}{1.000pt}}
\put(539,306){\rule[-0.500pt]{1.032pt}{1.000pt}}
\put(543,305){\usebox{\plotpoint}}
\put(548,304){\usebox{\plotpoint}}
\put(552,303){\usebox{\plotpoint}}
\put(556,302){\usebox{\plotpoint}}
\put(560,301){\usebox{\plotpoint}}
\put(564,300){\usebox{\plotpoint}}
\put(568,299){\usebox{\plotpoint}}
\put(573,298){\rule[-0.500pt]{1.032pt}{1.000pt}}
\put(577,297){\rule[-0.500pt]{1.032pt}{1.000pt}}
\put(581,296){\rule[-0.500pt]{1.032pt}{1.000pt}}
\put(585,295){\rule[-0.500pt]{1.032pt}{1.000pt}}
\put(590,294){\rule[-0.500pt]{1.032pt}{1.000pt}}
\put(594,293){\rule[-0.500pt]{1.032pt}{1.000pt}}
\put(598,292){\rule[-0.500pt]{1.032pt}{1.000pt}}
\put(602,291){\usebox{\plotpoint}}
\put(607,290){\usebox{\plotpoint}}
\put(611,289){\usebox{\plotpoint}}
\put(615,288){\usebox{\plotpoint}}
\put(619,287){\usebox{\plotpoint}}
\put(623,286){\usebox{\plotpoint}}
\put(627,285){\usebox{\plotpoint}}
\put(632,284){\rule[-0.500pt]{1.204pt}{1.000pt}}
\put(637,283){\rule[-0.500pt]{1.204pt}{1.000pt}}
\put(642,282){\rule[-0.500pt]{1.204pt}{1.000pt}}
\put(647,281){\rule[-0.500pt]{1.204pt}{1.000pt}}
\put(652,280){\rule[-0.500pt]{1.204pt}{1.000pt}}
\put(657,279){\rule[-0.500pt]{1.204pt}{1.000pt}}
\put(662,278){\rule[-0.500pt]{1.164pt}{1.000pt}}
\put(666,277){\rule[-0.500pt]{1.164pt}{1.000pt}}
\put(671,276){\rule[-0.500pt]{1.164pt}{1.000pt}}
\put(676,275){\rule[-0.500pt]{1.164pt}{1.000pt}}
\put(681,274){\rule[-0.500pt]{1.164pt}{1.000pt}}
\put(686,273){\rule[-0.500pt]{1.164pt}{1.000pt}}
\put(690,272){\rule[-0.500pt]{1.205pt}{1.000pt}}
\put(696,271){\rule[-0.500pt]{1.204pt}{1.000pt}}
\put(701,270){\rule[-0.500pt]{1.204pt}{1.000pt}}
\put(706,269){\rule[-0.500pt]{1.204pt}{1.000pt}}
\put(711,268){\rule[-0.500pt]{1.204pt}{1.000pt}}
\put(716,267){\rule[-0.500pt]{1.204pt}{1.000pt}}
\put(721,266){\rule[-0.500pt]{1.204pt}{1.000pt}}
\put(726,265){\rule[-0.500pt]{1.204pt}{1.000pt}}
\put(731,264){\rule[-0.500pt]{1.204pt}{1.000pt}}
\put(736,263){\rule[-0.500pt]{1.204pt}{1.000pt}}
\put(741,262){\rule[-0.500pt]{1.204pt}{1.000pt}}
\put(746,261){\rule[-0.500pt]{1.204pt}{1.000pt}}
\put(751,260){\rule[-0.500pt]{1.397pt}{1.000pt}}
\put(756,259){\rule[-0.500pt]{1.397pt}{1.000pt}}
\put(762,258){\rule[-0.500pt]{1.397pt}{1.000pt}}
\put(768,257){\rule[-0.500pt]{1.397pt}{1.000pt}}
\put(774,256){\rule[-0.500pt]{1.397pt}{1.000pt}}
\put(779,255){\rule[-0.500pt]{1.445pt}{1.000pt}}
\put(786,254){\rule[-0.500pt]{1.445pt}{1.000pt}}
\put(792,253){\rule[-0.500pt]{1.445pt}{1.000pt}}
\put(798,252){\rule[-0.500pt]{1.445pt}{1.000pt}}
\put(804,251){\rule[-0.500pt]{1.445pt}{1.000pt}}
\put(810,250){\rule[-0.500pt]{1.164pt}{1.000pt}}
\put(814,249){\rule[-0.500pt]{1.164pt}{1.000pt}}
\put(819,248){\rule[-0.500pt]{1.164pt}{1.000pt}}
\put(824,247){\rule[-0.500pt]{1.164pt}{1.000pt}}
\put(829,246){\rule[-0.500pt]{1.164pt}{1.000pt}}
\put(834,245){\rule[-0.500pt]{1.164pt}{1.000pt}}
\put(838,244){\rule[-0.500pt]{1.205pt}{1.000pt}}
\put(844,243){\rule[-0.500pt]{1.204pt}{1.000pt}}
\put(849,242){\rule[-0.500pt]{1.204pt}{1.000pt}}
\put(854,241){\rule[-0.500pt]{1.204pt}{1.000pt}}
\put(859,240){\rule[-0.500pt]{1.204pt}{1.000pt}}
\put(864,239){\rule[-0.500pt]{1.204pt}{1.000pt}}
\put(869,238){\rule[-0.500pt]{1.445pt}{1.000pt}}
\put(875,237){\rule[-0.500pt]{1.445pt}{1.000pt}}
\put(881,236){\rule[-0.500pt]{1.445pt}{1.000pt}}
\put(887,235){\rule[-0.500pt]{1.445pt}{1.000pt}}
\put(893,234){\rule[-0.500pt]{1.445pt}{1.000pt}}
\put(899,233){\rule[-0.500pt]{1.397pt}{1.000pt}}
\put(904,232){\rule[-0.500pt]{1.397pt}{1.000pt}}
\put(910,231){\rule[-0.500pt]{1.397pt}{1.000pt}}
\put(916,230){\rule[-0.500pt]{1.397pt}{1.000pt}}
\put(922,229){\rule[-0.500pt]{1.397pt}{1.000pt}}
\put(927,228){\rule[-0.500pt]{2.409pt}{1.000pt}}
\put(938,227){\rule[-0.500pt]{2.409pt}{1.000pt}}
\put(948,226){\rule[-0.500pt]{2.409pt}{1.000pt}}
\put(958,225){\rule[-0.500pt]{2.329pt}{1.000pt}}
\put(967,224){\rule[-0.500pt]{2.329pt}{1.000pt}}
\put(977,223){\rule[-0.500pt]{2.329pt}{1.000pt}}
\put(987,222){\rule[-0.500pt]{2.409pt}{1.000pt}}
\put(997,221){\rule[-0.500pt]{2.409pt}{1.000pt}}
\put(1007,220){\rule[-0.500pt]{2.409pt}{1.000pt}}
\put(1017,219){\rule[-0.500pt]{3.613pt}{1.000pt}}
\put(1032,218){\rule[-0.500pt]{3.613pt}{1.000pt}}
\put(1047,217){\rule[-0.500pt]{3.493pt}{1.000pt}}
\put(1061,216){\rule[-0.500pt]{3.493pt}{1.000pt}}
\put(1076,215){\rule[-0.500pt]{7.227pt}{1.000pt}}
\put(1106,214){\rule[-0.500pt]{6.986pt}{1.000pt}}
\put(1135,213){\rule[-0.500pt]{7.227pt}{1.000pt}}
\put(1165,212){\rule[-0.500pt]{7.227pt}{1.000pt}}
\put(1195,211){\rule[-0.500pt]{6.986pt}{1.000pt}}
\put(1224,210){\rule[-0.500pt]{35.653pt}{1.000pt}}
\put(1306,677){\makebox(0,0)[r]{{\ft light higgs exp}}}
\put(1328,677){\rule[-0.500pt]{15.899pt}{1.000pt}}
\put(270,459){\usebox{\plotpoint}}
\put(270,459){\usebox{\plotpoint}}
\put(271,458){\usebox{\plotpoint}}
\put(272,457){\usebox{\plotpoint}}
\put(273,456){\usebox{\plotpoint}}
\put(275,455){\usebox{\plotpoint}}
\put(276,454){\usebox{\plotpoint}}
\put(277,453){\usebox{\plotpoint}}
\put(278,452){\usebox{\plotpoint}}
\put(280,451){\usebox{\plotpoint}}
\put(281,450){\usebox{\plotpoint}}
\put(282,449){\usebox{\plotpoint}}
\put(283,448){\usebox{\plotpoint}}
\put(284,447){\usebox{\plotpoint}}
\put(285,446){\usebox{\plotpoint}}
\put(286,445){\usebox{\plotpoint}}
\put(288,444){\usebox{\plotpoint}}
\put(289,443){\usebox{\plotpoint}}
\put(290,442){\usebox{\plotpoint}}
\put(292,441){\usebox{\plotpoint}}
\put(293,440){\usebox{\plotpoint}}
\put(294,439){\usebox{\plotpoint}}
\put(295,438){\usebox{\plotpoint}}
\put(297,437){\usebox{\plotpoint}}
\put(299,436){\usebox{\plotpoint}}
\put(300,435){\usebox{\plotpoint}}
\put(302,434){\usebox{\plotpoint}}
\put(303,433){\usebox{\plotpoint}}
\put(305,432){\usebox{\plotpoint}}
\put(306,431){\usebox{\plotpoint}}
\put(308,430){\usebox{\plotpoint}}
\put(309,429){\usebox{\plotpoint}}
\put(311,428){\usebox{\plotpoint}}
\put(312,427){\usebox{\plotpoint}}
\put(314,426){\usebox{\plotpoint}}
\put(315,425){\usebox{\plotpoint}}
\put(317,424){\usebox{\plotpoint}}
\put(318,423){\usebox{\plotpoint}}
\put(320,422){\usebox{\plotpoint}}
\put(321,421){\usebox{\plotpoint}}
\put(323,420){\usebox{\plotpoint}}
\put(324,419){\usebox{\plotpoint}}
\put(325,418){\usebox{\plotpoint}}
\put(327,417){\usebox{\plotpoint}}
\put(328,416){\usebox{\plotpoint}}
\put(329,415){\usebox{\plotpoint}}
\put(331,414){\usebox{\plotpoint}}
\put(332,413){\usebox{\plotpoint}}
\put(334,412){\usebox{\plotpoint}}
\put(335,411){\usebox{\plotpoint}}
\put(337,410){\usebox{\plotpoint}}
\put(338,409){\usebox{\plotpoint}}
\put(340,408){\usebox{\plotpoint}}
\put(341,407){\usebox{\plotpoint}}
\put(343,406){\usebox{\plotpoint}}
\put(344,405){\usebox{\plotpoint}}
\put(346,404){\usebox{\plotpoint}}
\put(347,403){\usebox{\plotpoint}}
\put(349,402){\usebox{\plotpoint}}
\put(350,401){\usebox{\plotpoint}}
\put(352,400){\usebox{\plotpoint}}
\put(354,399){\usebox{\plotpoint}}
\put(356,398){\usebox{\plotpoint}}
\put(357,397){\usebox{\plotpoint}}
\put(359,396){\usebox{\plotpoint}}
\put(361,395){\usebox{\plotpoint}}
\put(363,394){\usebox{\plotpoint}}
\put(365,393){\usebox{\plotpoint}}
\put(366,392){\usebox{\plotpoint}}
\put(368,391){\usebox{\plotpoint}}
\put(370,390){\usebox{\plotpoint}}
\put(371,389){\usebox{\plotpoint}}
\put(373,388){\usebox{\plotpoint}}
\put(375,387){\usebox{\plotpoint}}
\put(376,386){\usebox{\plotpoint}}
\put(378,385){\usebox{\plotpoint}}
\put(380,384){\usebox{\plotpoint}}
\put(382,383){\usebox{\plotpoint}}
\put(383,382){\usebox{\plotpoint}}
\put(385,381){\usebox{\plotpoint}}
\put(387,380){\usebox{\plotpoint}}
\put(389,379){\usebox{\plotpoint}}
\put(391,378){\usebox{\plotpoint}}
\put(392,377){\usebox{\plotpoint}}
\put(395,376){\usebox{\plotpoint}}
\put(397,375){\usebox{\plotpoint}}
\put(399,374){\usebox{\plotpoint}}
\put(402,373){\usebox{\plotpoint}}
\put(404,372){\usebox{\plotpoint}}
\put(406,371){\usebox{\plotpoint}}
\put(408,370){\usebox{\plotpoint}}
\put(411,369){\usebox{\plotpoint}}
\put(413,368){\usebox{\plotpoint}}
\put(415,367){\usebox{\plotpoint}}
\put(417,366){\usebox{\plotpoint}}
\put(419,365){\usebox{\plotpoint}}
\put(421,364){\usebox{\plotpoint}}
\put(423,363){\usebox{\plotpoint}}
\put(425,362){\usebox{\plotpoint}}
\put(428,361){\usebox{\plotpoint}}
\put(431,360){\usebox{\plotpoint}}
\put(434,359){\usebox{\plotpoint}}
\put(437,358){\usebox{\plotpoint}}
\put(440,357){\usebox{\plotpoint}}
\put(443,356){\usebox{\plotpoint}}
\put(446,355){\usebox{\plotpoint}}
\put(448,354){\usebox{\plotpoint}}
\put(451,353){\usebox{\plotpoint}}
\put(453,352){\usebox{\plotpoint}}
\put(457,351){\usebox{\plotpoint}}
\put(460,350){\usebox{\plotpoint}}
\put(463,349){\rule[-0.500pt]{1.084pt}{1.000pt}}
\put(467,348){\rule[-0.500pt]{1.084pt}{1.000pt}}
\put(472,347){\rule[-0.500pt]{1.084pt}{1.000pt}}
\put(476,346){\rule[-0.500pt]{1.084pt}{1.000pt}}
\put(481,345){\rule[-0.500pt]{2.168pt}{1.000pt}}
\put(490,344){\usebox{\plotpoint}}
\put(494,343){\usebox{\plotpoint}}
\put(498,342){\rule[-0.500pt]{2.168pt}{1.000pt}}
\sbox{\plotpoint}{\rule[-0.350pt]{0.700pt}{0.700pt}}%
\put(1306,632){\makebox(0,0)[r]{{\ft interpolation}}}
\put(1328,632){\rule[-0.350pt]{15.899pt}{0.700pt}}
\put(270,459){\usebox{\plotpoint}}
\put(270,459){\usebox{\plotpoint}}
\put(271,458){\usebox{\plotpoint}}
\put(272,457){\usebox{\plotpoint}}
\put(273,456){\usebox{\plotpoint}}
\put(275,455){\usebox{\plotpoint}}
\put(276,454){\usebox{\plotpoint}}
\put(277,453){\usebox{\plotpoint}}
\put(278,452){\usebox{\plotpoint}}
\put(280,451){\usebox{\plotpoint}}
\put(281,450){\usebox{\plotpoint}}
\put(282,449){\usebox{\plotpoint}}
\put(283,448){\usebox{\plotpoint}}
\put(285,447){\usebox{\plotpoint}}
\put(286,446){\usebox{\plotpoint}}
\put(287,445){\usebox{\plotpoint}}
\put(288,444){\usebox{\plotpoint}}
\put(290,443){\usebox{\plotpoint}}
\put(291,442){\usebox{\plotpoint}}
\put(292,441){\usebox{\plotpoint}}
\put(293,440){\usebox{\plotpoint}}
\put(295,439){\usebox{\plotpoint}}
\put(296,438){\usebox{\plotpoint}}
\put(297,437){\usebox{\plotpoint}}
\put(299,436){\usebox{\plotpoint}}
\put(300,435){\usebox{\plotpoint}}
\put(302,434){\usebox{\plotpoint}}
\put(303,433){\usebox{\plotpoint}}
\put(305,432){\usebox{\plotpoint}}
\put(306,431){\usebox{\plotpoint}}
\put(308,430){\usebox{\plotpoint}}
\put(309,429){\usebox{\plotpoint}}
\put(311,428){\usebox{\plotpoint}}
\put(312,427){\usebox{\plotpoint}}
\put(314,426){\usebox{\plotpoint}}
\put(315,425){\usebox{\plotpoint}}
\put(316,424){\usebox{\plotpoint}}
\put(318,423){\usebox{\plotpoint}}
\put(319,422){\usebox{\plotpoint}}
\put(320,421){\usebox{\plotpoint}}
\put(322,420){\usebox{\plotpoint}}
\put(323,419){\usebox{\plotpoint}}
\put(325,418){\usebox{\plotpoint}}
\put(326,417){\usebox{\plotpoint}}
\put(327,416){\usebox{\plotpoint}}
\put(329,415){\usebox{\plotpoint}}
\put(331,414){\usebox{\plotpoint}}
\put(332,413){\usebox{\plotpoint}}
\put(334,412){\usebox{\plotpoint}}
\put(335,411){\usebox{\plotpoint}}
\put(337,410){\usebox{\plotpoint}}
\put(338,409){\usebox{\plotpoint}}
\put(340,408){\usebox{\plotpoint}}
\put(341,407){\usebox{\plotpoint}}
\put(343,406){\usebox{\plotpoint}}
\put(344,405){\usebox{\plotpoint}}
\put(346,404){\usebox{\plotpoint}}
\put(347,403){\usebox{\plotpoint}}
\put(349,402){\usebox{\plotpoint}}
\put(351,401){\usebox{\plotpoint}}
\put(352,400){\usebox{\plotpoint}}
\put(354,399){\usebox{\plotpoint}}
\put(356,398){\usebox{\plotpoint}}
\put(357,397){\usebox{\plotpoint}}
\put(359,396){\usebox{\plotpoint}}
\put(361,395){\usebox{\plotpoint}}
\put(362,394){\usebox{\plotpoint}}
\put(364,393){\usebox{\plotpoint}}
\put(365,392){\usebox{\plotpoint}}
\put(367,391){\usebox{\plotpoint}}
\put(368,390){\usebox{\plotpoint}}
\put(370,389){\usebox{\plotpoint}}
\put(371,388){\usebox{\plotpoint}}
\put(373,387){\usebox{\plotpoint}}
\put(375,386){\usebox{\plotpoint}}
\put(377,385){\usebox{\plotpoint}}
\put(379,384){\usebox{\plotpoint}}
\put(381,383){\usebox{\plotpoint}}
\put(383,382){\usebox{\plotpoint}}
\put(385,381){\usebox{\plotpoint}}
\put(387,380){\usebox{\plotpoint}}
\put(388,379){\usebox{\plotpoint}}
\put(390,378){\usebox{\plotpoint}}
\put(392,377){\usebox{\plotpoint}}
\put(394,376){\usebox{\plotpoint}}
\put(396,375){\usebox{\plotpoint}}
\put(398,374){\usebox{\plotpoint}}
\put(400,373){\usebox{\plotpoint}}
\put(402,372){\usebox{\plotpoint}}
\put(403,371){\usebox{\plotpoint}}
\put(405,370){\usebox{\plotpoint}}
\put(407,369){\usebox{\plotpoint}}
\put(409,368){\usebox{\plotpoint}}
\put(410,367){\usebox{\plotpoint}}
\put(412,366){\usebox{\plotpoint}}
\put(414,365){\usebox{\plotpoint}}
\put(416,364){\usebox{\plotpoint}}
\put(418,363){\usebox{\plotpoint}}
\put(420,362){\usebox{\plotpoint}}
\put(422,361){\usebox{\plotpoint}}
\put(424,360){\usebox{\plotpoint}}
\put(426,359){\usebox{\plotpoint}}
\put(428,358){\usebox{\plotpoint}}
\put(430,357){\usebox{\plotpoint}}
\put(433,356){\usebox{\plotpoint}}
\put(435,355){\usebox{\plotpoint}}
\put(437,354){\usebox{\plotpoint}}
\put(439,353){\usebox{\plotpoint}}
\put(441,352){\usebox{\plotpoint}}
\put(443,351){\usebox{\plotpoint}}
\put(445,350){\usebox{\plotpoint}}
\put(448,349){\usebox{\plotpoint}}
\put(450,348){\usebox{\plotpoint}}
\put(452,347){\usebox{\plotpoint}}
\put(454,346){\usebox{\plotpoint}}
\put(456,345){\usebox{\plotpoint}}
\put(458,344){\usebox{\plotpoint}}
\put(460,343){\usebox{\plotpoint}}
\put(462,342){\usebox{\plotpoint}}
\put(465,341){\usebox{\plotpoint}}
\put(467,340){\usebox{\plotpoint}}
\put(470,339){\usebox{\plotpoint}}
\put(472,338){\usebox{\plotpoint}}
\put(475,337){\usebox{\plotpoint}}
\put(477,336){\usebox{\plotpoint}}
\put(480,335){\usebox{\plotpoint}}
\put(482,334){\usebox{\plotpoint}}
\put(485,333){\usebox{\plotpoint}}
\put(487,332){\usebox{\plotpoint}}
\put(490,331){\usebox{\plotpoint}}
\put(492,330){\usebox{\plotpoint}}
\put(494,329){\usebox{\plotpoint}}
\put(497,328){\usebox{\plotpoint}}
\put(499,327){\usebox{\plotpoint}}
\put(501,326){\usebox{\plotpoint}}
\put(504,325){\usebox{\plotpoint}}
\put(506,324){\usebox{\plotpoint}}
\put(509,323){\usebox{\plotpoint}}
\put(511,322){\usebox{\plotpoint}}
\put(514,321){\usebox{\plotpoint}}
\put(516,320){\usebox{\plotpoint}}
\put(519,319){\rule[-0.350pt]{0.723pt}{0.700pt}}
\put(522,318){\rule[-0.350pt]{0.723pt}{0.700pt}}
\put(525,317){\rule[-0.350pt]{0.723pt}{0.700pt}}
\put(528,316){\rule[-0.350pt]{0.723pt}{0.700pt}}
\put(531,315){\rule[-0.350pt]{0.723pt}{0.700pt}}
\put(534,314){\usebox{\plotpoint}}
\put(536,313){\usebox{\plotpoint}}
\put(539,312){\usebox{\plotpoint}}
\put(542,311){\usebox{\plotpoint}}
\put(545,310){\usebox{\plotpoint}}
\put(547,309){\rule[-0.350pt]{0.723pt}{0.700pt}}
\put(551,308){\rule[-0.350pt]{0.723pt}{0.700pt}}
\put(554,307){\rule[-0.350pt]{0.723pt}{0.700pt}}
\put(557,306){\rule[-0.350pt]{0.723pt}{0.700pt}}
\put(560,305){\rule[-0.350pt]{0.723pt}{0.700pt}}
\put(563,304){\rule[-0.350pt]{0.903pt}{0.700pt}}
\put(566,303){\rule[-0.350pt]{0.903pt}{0.700pt}}
\put(570,302){\rule[-0.350pt]{0.903pt}{0.700pt}}
\put(574,301){\rule[-0.350pt]{0.903pt}{0.700pt}}
\put(578,300){\usebox{\plotpoint}}
\put(580,299){\usebox{\plotpoint}}
\put(583,298){\usebox{\plotpoint}}
\put(586,297){\usebox{\plotpoint}}
\put(589,296){\usebox{\plotpoint}}
\put(591,295){\rule[-0.350pt]{0.903pt}{0.700pt}}
\put(595,294){\rule[-0.350pt]{0.903pt}{0.700pt}}
\put(599,293){\rule[-0.350pt]{0.903pt}{0.700pt}}
\put(603,292){\rule[-0.350pt]{0.903pt}{0.700pt}}
\put(607,291){\rule[-0.350pt]{0.843pt}{0.700pt}}
\put(610,290){\rule[-0.350pt]{0.843pt}{0.700pt}}
\put(614,289){\rule[-0.350pt]{0.843pt}{0.700pt}}
\put(617,288){\rule[-0.350pt]{0.843pt}{0.700pt}}
\put(621,287){\rule[-0.350pt]{0.903pt}{0.700pt}}
\put(624,286){\rule[-0.350pt]{0.903pt}{0.700pt}}
\put(628,285){\rule[-0.350pt]{0.903pt}{0.700pt}}
\put(632,284){\rule[-0.350pt]{0.903pt}{0.700pt}}
\put(636,283){\rule[-0.350pt]{0.903pt}{0.700pt}}
\put(639,282){\rule[-0.350pt]{0.903pt}{0.700pt}}
\put(643,281){\rule[-0.350pt]{0.903pt}{0.700pt}}
\put(647,280){\rule[-0.350pt]{0.903pt}{0.700pt}}
\put(651,279){\rule[-0.350pt]{1.124pt}{0.700pt}}
\put(655,278){\rule[-0.350pt]{1.124pt}{0.700pt}}
\put(660,277){\rule[-0.350pt]{1.124pt}{0.700pt}}
\put(665,276){\rule[-0.350pt]{0.903pt}{0.700pt}}
\put(668,275){\rule[-0.350pt]{0.903pt}{0.700pt}}
\put(672,274){\rule[-0.350pt]{0.903pt}{0.700pt}}
\put(676,273){\rule[-0.350pt]{0.903pt}{0.700pt}}
\put(680,272){\rule[-0.350pt]{1.204pt}{0.700pt}}
\put(685,271){\rule[-0.350pt]{1.204pt}{0.700pt}}
\put(690,270){\rule[-0.350pt]{1.204pt}{0.700pt}}
\put(695,269){\rule[-0.350pt]{1.124pt}{0.700pt}}
\put(699,268){\rule[-0.350pt]{1.124pt}{0.700pt}}
\put(704,267){\rule[-0.350pt]{1.124pt}{0.700pt}}
\put(709,266){\rule[-0.350pt]{1.204pt}{0.700pt}}
\put(714,265){\rule[-0.350pt]{1.204pt}{0.700pt}}
\put(719,264){\rule[-0.350pt]{1.204pt}{0.700pt}}
\put(724,263){\rule[-0.350pt]{1.204pt}{0.700pt}}
\put(729,262){\rule[-0.350pt]{1.204pt}{0.700pt}}
\put(734,261){\rule[-0.350pt]{1.204pt}{0.700pt}}
\put(739,260){\rule[-0.350pt]{1.686pt}{0.700pt}}
\put(746,259){\rule[-0.350pt]{1.686pt}{0.700pt}}
\put(753,258){\rule[-0.350pt]{1.204pt}{0.700pt}}
\put(758,257){\rule[-0.350pt]{1.204pt}{0.700pt}}
\put(763,256){\rule[-0.350pt]{1.204pt}{0.700pt}}
\put(768,255){\rule[-0.350pt]{1.204pt}{0.700pt}}
\put(773,254){\rule[-0.350pt]{1.204pt}{0.700pt}}
\put(778,253){\rule[-0.350pt]{1.204pt}{0.700pt}}
\put(783,252){\rule[-0.350pt]{1.686pt}{0.700pt}}
\put(790,251){\rule[-0.350pt]{1.686pt}{0.700pt}}
\put(797,250){\rule[-0.350pt]{1.204pt}{0.700pt}}
\put(802,249){\rule[-0.350pt]{1.204pt}{0.700pt}}
\put(807,248){\rule[-0.350pt]{1.204pt}{0.700pt}}
\put(812,247){\rule[-0.350pt]{1.807pt}{0.700pt}}
\put(819,246){\rule[-0.350pt]{1.807pt}{0.700pt}}
\put(827,245){\rule[-0.350pt]{1.686pt}{0.700pt}}
\put(834,244){\rule[-0.350pt]{1.686pt}{0.700pt}}
\put(841,243){\rule[-0.350pt]{1.807pt}{0.700pt}}
\put(848,242){\rule[-0.350pt]{1.807pt}{0.700pt}}
\put(856,241){\rule[-0.350pt]{1.807pt}{0.700pt}}
\put(863,240){\rule[-0.350pt]{1.807pt}{0.700pt}}
\put(871,239){\rule[-0.350pt]{1.686pt}{0.700pt}}
\put(878,238){\rule[-0.350pt]{1.686pt}{0.700pt}}
\put(885,237){\rule[-0.350pt]{1.807pt}{0.700pt}}
\put(892,236){\rule[-0.350pt]{1.807pt}{0.700pt}}
\put(900,235){\rule[-0.350pt]{1.686pt}{0.700pt}}
\put(907,234){\rule[-0.350pt]{1.686pt}{0.700pt}}
\put(914,233){\rule[-0.350pt]{1.807pt}{0.700pt}}
\put(921,232){\rule[-0.350pt]{1.807pt}{0.700pt}}
\put(929,231){\rule[-0.350pt]{3.613pt}{0.700pt}}
\put(944,230){\rule[-0.350pt]{1.686pt}{0.700pt}}
\put(951,229){\rule[-0.350pt]{1.686pt}{0.700pt}}
\put(958,228){\rule[-0.350pt]{1.807pt}{0.700pt}}
\put(965,227){\rule[-0.350pt]{1.807pt}{0.700pt}}
\put(973,226){\rule[-0.350pt]{3.613pt}{0.700pt}}
\put(988,225){\rule[-0.350pt]{1.686pt}{0.700pt}}
\put(995,224){\rule[-0.350pt]{1.686pt}{0.700pt}}
\put(1002,223){\rule[-0.350pt]{3.613pt}{0.700pt}}
\put(1017,222){\rule[-0.350pt]{3.613pt}{0.700pt}}
\put(1032,221){\rule[-0.350pt]{1.686pt}{0.700pt}}
\put(1039,220){\rule[-0.350pt]{1.686pt}{0.700pt}}
\put(1046,219){\rule[-0.350pt]{3.613pt}{0.700pt}}
\put(1061,218){\rule[-0.350pt]{3.613pt}{0.700pt}}
\put(1076,217){\rule[-0.350pt]{3.373pt}{0.700pt}}
\put(1090,216){\rule[-0.350pt]{3.613pt}{0.700pt}}
\put(1105,215){\rule[-0.350pt]{3.613pt}{0.700pt}}
\put(1120,214){\rule[-0.350pt]{3.373pt}{0.700pt}}
\put(1134,213){\rule[-0.350pt]{3.613pt}{0.700pt}}
\put(1149,212){\rule[-0.350pt]{3.613pt}{0.700pt}}
\put(1164,211){\rule[-0.350pt]{6.986pt}{0.700pt}}
\put(1193,210){\rule[-0.350pt]{10.600pt}{0.700pt}}
\put(1237,209){\rule[-0.350pt]{13.972pt}{0.700pt}}
\put(1295,210){\rule[-0.350pt]{7.227pt}{0.700pt}}
\put(1325,211){\rule[-0.350pt]{3.373pt}{0.700pt}}
\put(1339,212){\rule[-0.350pt]{5.420pt}{0.700pt}}
\put(1361,213){\rule[-0.350pt]{1.807pt}{0.700pt}}
\sbox{\plotpoint}{\rule[-0.175pt]{0.350pt}{0.350pt}}%
\put(1306,587){\makebox(0,0)[r]{{\ft leading $m_t^4$}}}
\put(1328,587){\rule[-0.175pt]{15.899pt}{0.350pt}}
\put(270,753){\usebox{\plotpoint}}
\put(270,751){\usebox{\plotpoint}}
\put(271,750){\usebox{\plotpoint}}
\put(272,749){\usebox{\plotpoint}}
\put(273,748){\usebox{\plotpoint}}
\put(274,746){\usebox{\plotpoint}}
\put(275,745){\usebox{\plotpoint}}
\put(276,744){\usebox{\plotpoint}}
\put(277,743){\usebox{\plotpoint}}
\put(278,741){\usebox{\plotpoint}}
\put(279,740){\usebox{\plotpoint}}
\put(280,739){\usebox{\plotpoint}}
\put(281,738){\usebox{\plotpoint}}
\put(282,736){\usebox{\plotpoint}}
\put(283,735){\usebox{\plotpoint}}
\put(284,734){\usebox{\plotpoint}}
\put(285,733){\usebox{\plotpoint}}
\put(286,731){\usebox{\plotpoint}}
\put(287,730){\usebox{\plotpoint}}
\put(288,729){\usebox{\plotpoint}}
\put(289,728){\usebox{\plotpoint}}
\put(290,726){\usebox{\plotpoint}}
\put(291,725){\usebox{\plotpoint}}
\put(292,724){\usebox{\plotpoint}}
\put(293,723){\usebox{\plotpoint}}
\put(294,721){\usebox{\plotpoint}}
\put(295,720){\usebox{\plotpoint}}
\put(296,719){\usebox{\plotpoint}}
\put(297,718){\usebox{\plotpoint}}
\put(298,717){\usebox{\plotpoint}}
\put(299,715){\usebox{\plotpoint}}
\put(300,714){\usebox{\plotpoint}}
\put(301,713){\usebox{\plotpoint}}
\put(302,712){\usebox{\plotpoint}}
\put(303,711){\usebox{\plotpoint}}
\put(304,710){\usebox{\plotpoint}}
\put(305,709){\usebox{\plotpoint}}
\put(306,708){\usebox{\plotpoint}}
\put(307,707){\usebox{\plotpoint}}
\put(308,706){\usebox{\plotpoint}}
\put(309,705){\usebox{\plotpoint}}
\put(310,704){\usebox{\plotpoint}}
\put(311,703){\usebox{\plotpoint}}
\put(312,702){\usebox{\plotpoint}}
\put(313,701){\usebox{\plotpoint}}
\put(314,700){\usebox{\plotpoint}}
\put(315,699){\usebox{\plotpoint}}
\put(316,698){\usebox{\plotpoint}}
\put(317,697){\usebox{\plotpoint}}
\put(318,696){\usebox{\plotpoint}}
\put(319,695){\usebox{\plotpoint}}
\put(320,694){\usebox{\plotpoint}}
\put(321,693){\usebox{\plotpoint}}
\put(322,692){\usebox{\plotpoint}}
\put(323,691){\usebox{\plotpoint}}
\put(324,690){\usebox{\plotpoint}}
\put(325,689){\usebox{\plotpoint}}
\put(326,688){\usebox{\plotpoint}}
\put(327,687){\usebox{\plotpoint}}
\put(328,687){\usebox{\plotpoint}}
\put(329,686){\usebox{\plotpoint}}
\put(330,685){\usebox{\plotpoint}}
\put(331,684){\usebox{\plotpoint}}
\put(332,683){\usebox{\plotpoint}}
\put(334,682){\usebox{\plotpoint}}
\put(335,681){\usebox{\plotpoint}}
\put(336,680){\usebox{\plotpoint}}
\put(337,679){\usebox{\plotpoint}}
\put(338,678){\usebox{\plotpoint}}
\put(340,677){\usebox{\plotpoint}}
\put(341,676){\usebox{\plotpoint}}
\put(342,675){\usebox{\plotpoint}}
\put(343,674){\usebox{\plotpoint}}
\put(344,673){\usebox{\plotpoint}}
\put(346,672){\usebox{\plotpoint}}
\put(347,671){\usebox{\plotpoint}}
\put(348,670){\usebox{\plotpoint}}
\put(349,669){\usebox{\plotpoint}}
\put(350,668){\usebox{\plotpoint}}
\put(352,667){\usebox{\plotpoint}}
\put(353,666){\usebox{\plotpoint}}
\put(354,665){\usebox{\plotpoint}}
\put(355,664){\usebox{\plotpoint}}
\put(356,663){\usebox{\plotpoint}}
\put(358,662){\usebox{\plotpoint}}
\put(359,661){\usebox{\plotpoint}}
\put(360,660){\usebox{\plotpoint}}
\put(361,659){\usebox{\plotpoint}}
\put(363,658){\usebox{\plotpoint}}
\put(364,657){\usebox{\plotpoint}}
\put(365,656){\usebox{\plotpoint}}
\put(366,655){\usebox{\plotpoint}}
\put(368,654){\usebox{\plotpoint}}
\put(369,653){\usebox{\plotpoint}}
\put(370,652){\usebox{\plotpoint}}
\put(371,651){\usebox{\plotpoint}}
\put(373,650){\usebox{\plotpoint}}
\put(374,649){\usebox{\plotpoint}}
\put(375,648){\usebox{\plotpoint}}
\put(376,647){\usebox{\plotpoint}}
\put(378,646){\usebox{\plotpoint}}
\put(379,645){\usebox{\plotpoint}}
\put(380,644){\usebox{\plotpoint}}
\put(381,643){\usebox{\plotpoint}}
\put(383,642){\usebox{\plotpoint}}
\put(384,641){\usebox{\plotpoint}}
\put(385,640){\usebox{\plotpoint}}
\put(386,639){\usebox{\plotpoint}}
\put(388,638){\usebox{\plotpoint}}
\put(389,637){\usebox{\plotpoint}}
\put(391,636){\usebox{\plotpoint}}
\put(392,635){\usebox{\plotpoint}}
\put(394,634){\usebox{\plotpoint}}
\put(395,633){\usebox{\plotpoint}}
\put(397,632){\usebox{\plotpoint}}
\put(398,631){\usebox{\plotpoint}}
\put(400,630){\usebox{\plotpoint}}
\put(401,629){\usebox{\plotpoint}}
\put(402,628){\usebox{\plotpoint}}
\put(404,627){\usebox{\plotpoint}}
\put(405,626){\usebox{\plotpoint}}
\put(407,625){\usebox{\plotpoint}}
\put(408,624){\usebox{\plotpoint}}
\put(410,623){\usebox{\plotpoint}}
\put(411,622){\usebox{\plotpoint}}
\put(413,621){\usebox{\plotpoint}}
\put(414,620){\usebox{\plotpoint}}
\put(416,619){\rule[-0.175pt]{0.380pt}{0.350pt}}
\put(417,618){\rule[-0.175pt]{0.380pt}{0.350pt}}
\put(419,617){\rule[-0.175pt]{0.380pt}{0.350pt}}
\put(420,616){\rule[-0.175pt]{0.380pt}{0.350pt}}
\put(422,615){\rule[-0.175pt]{0.380pt}{0.350pt}}
\put(423,614){\rule[-0.175pt]{0.380pt}{0.350pt}}
\put(425,613){\rule[-0.175pt]{0.380pt}{0.350pt}}
\put(427,612){\rule[-0.175pt]{0.380pt}{0.350pt}}
\put(428,611){\rule[-0.175pt]{0.380pt}{0.350pt}}
\put(430,610){\rule[-0.175pt]{0.380pt}{0.350pt}}
\put(431,609){\rule[-0.175pt]{0.380pt}{0.350pt}}
\put(433,608){\rule[-0.175pt]{0.380pt}{0.350pt}}
\put(434,607){\rule[-0.175pt]{0.380pt}{0.350pt}}
\put(436,606){\rule[-0.175pt]{0.380pt}{0.350pt}}
\put(438,605){\rule[-0.175pt]{0.380pt}{0.350pt}}
\put(439,604){\rule[-0.175pt]{0.380pt}{0.350pt}}
\put(441,603){\rule[-0.175pt]{0.380pt}{0.350pt}}
\put(442,602){\rule[-0.175pt]{0.380pt}{0.350pt}}
\put(444,601){\rule[-0.175pt]{0.380pt}{0.350pt}}
\put(446,600){\rule[-0.175pt]{0.437pt}{0.350pt}}
\put(447,599){\rule[-0.175pt]{0.437pt}{0.350pt}}
\put(449,598){\rule[-0.175pt]{0.437pt}{0.350pt}}
\put(451,597){\rule[-0.175pt]{0.437pt}{0.350pt}}
\put(453,596){\rule[-0.175pt]{0.437pt}{0.350pt}}
\put(455,595){\rule[-0.175pt]{0.437pt}{0.350pt}}
\put(456,594){\rule[-0.175pt]{0.437pt}{0.350pt}}
\put(458,593){\rule[-0.175pt]{0.437pt}{0.350pt}}
\put(460,592){\rule[-0.175pt]{0.437pt}{0.350pt}}
\put(462,591){\rule[-0.175pt]{0.437pt}{0.350pt}}
\put(464,590){\rule[-0.175pt]{0.437pt}{0.350pt}}
\put(465,589){\rule[-0.175pt]{0.437pt}{0.350pt}}
\put(467,588){\rule[-0.175pt]{0.437pt}{0.350pt}}
\put(469,587){\rule[-0.175pt]{0.437pt}{0.350pt}}
\put(471,586){\rule[-0.175pt]{0.437pt}{0.350pt}}
\put(473,585){\rule[-0.175pt]{0.437pt}{0.350pt}}
\put(475,584){\rule[-0.175pt]{0.466pt}{0.350pt}}
\put(476,583){\rule[-0.175pt]{0.466pt}{0.350pt}}
\put(478,582){\rule[-0.175pt]{0.466pt}{0.350pt}}
\put(480,581){\rule[-0.175pt]{0.466pt}{0.350pt}}
\put(482,580){\rule[-0.175pt]{0.466pt}{0.350pt}}
\put(484,579){\rule[-0.175pt]{0.466pt}{0.350pt}}
\put(486,578){\rule[-0.175pt]{0.466pt}{0.350pt}}
\put(488,577){\rule[-0.175pt]{0.466pt}{0.350pt}}
\put(490,576){\rule[-0.175pt]{0.466pt}{0.350pt}}
\put(492,575){\rule[-0.175pt]{0.466pt}{0.350pt}}
\put(494,574){\rule[-0.175pt]{0.466pt}{0.350pt}}
\put(496,573){\rule[-0.175pt]{0.466pt}{0.350pt}}
\put(498,572){\rule[-0.175pt]{0.466pt}{0.350pt}}
\put(500,571){\rule[-0.175pt]{0.466pt}{0.350pt}}
\put(502,570){\rule[-0.175pt]{0.466pt}{0.350pt}}
\put(503,569){\rule[-0.175pt]{0.516pt}{0.350pt}}
\put(506,568){\rule[-0.175pt]{0.516pt}{0.350pt}}
\put(508,567){\rule[-0.175pt]{0.516pt}{0.350pt}}
\put(510,566){\rule[-0.175pt]{0.516pt}{0.350pt}}
\put(512,565){\rule[-0.175pt]{0.516pt}{0.350pt}}
\put(514,564){\rule[-0.175pt]{0.516pt}{0.350pt}}
\put(516,563){\rule[-0.175pt]{0.516pt}{0.350pt}}
\put(519,562){\rule[-0.175pt]{0.516pt}{0.350pt}}
\put(521,561){\rule[-0.175pt]{0.516pt}{0.350pt}}
\put(523,560){\rule[-0.175pt]{0.516pt}{0.350pt}}
\put(525,559){\rule[-0.175pt]{0.516pt}{0.350pt}}
\put(527,558){\rule[-0.175pt]{0.516pt}{0.350pt}}
\put(529,557){\rule[-0.175pt]{0.516pt}{0.350pt}}
\put(531,556){\rule[-0.175pt]{0.516pt}{0.350pt}}
\put(534,555){\rule[-0.175pt]{0.537pt}{0.350pt}}
\put(536,554){\rule[-0.175pt]{0.537pt}{0.350pt}}
\put(538,553){\rule[-0.175pt]{0.537pt}{0.350pt}}
\put(540,552){\rule[-0.175pt]{0.537pt}{0.350pt}}
\put(542,551){\rule[-0.175pt]{0.537pt}{0.350pt}}
\put(545,550){\rule[-0.175pt]{0.537pt}{0.350pt}}
\put(547,549){\rule[-0.175pt]{0.537pt}{0.350pt}}
\put(549,548){\rule[-0.175pt]{0.537pt}{0.350pt}}
\put(551,547){\rule[-0.175pt]{0.537pt}{0.350pt}}
\put(554,546){\rule[-0.175pt]{0.537pt}{0.350pt}}
\put(556,545){\rule[-0.175pt]{0.537pt}{0.350pt}}
\put(558,544){\rule[-0.175pt]{0.537pt}{0.350pt}}
\put(560,543){\rule[-0.175pt]{0.537pt}{0.350pt}}
\put(563,542){\rule[-0.175pt]{0.582pt}{0.350pt}}
\put(565,541){\rule[-0.175pt]{0.582pt}{0.350pt}}
\put(567,540){\rule[-0.175pt]{0.582pt}{0.350pt}}
\put(570,539){\rule[-0.175pt]{0.582pt}{0.350pt}}
\put(572,538){\rule[-0.175pt]{0.582pt}{0.350pt}}
\put(575,537){\rule[-0.175pt]{0.582pt}{0.350pt}}
\put(577,536){\rule[-0.175pt]{0.582pt}{0.350pt}}
\put(579,535){\rule[-0.175pt]{0.582pt}{0.350pt}}
\put(582,534){\rule[-0.175pt]{0.582pt}{0.350pt}}
\put(584,533){\rule[-0.175pt]{0.582pt}{0.350pt}}
\put(587,532){\rule[-0.175pt]{0.582pt}{0.350pt}}
\put(589,531){\rule[-0.175pt]{0.582pt}{0.350pt}}
\put(592,530){\rule[-0.175pt]{0.699pt}{0.350pt}}
\put(594,529){\rule[-0.175pt]{0.699pt}{0.350pt}}
\put(597,528){\rule[-0.175pt]{0.699pt}{0.350pt}}
\put(600,527){\rule[-0.175pt]{0.699pt}{0.350pt}}
\put(603,526){\rule[-0.175pt]{0.699pt}{0.350pt}}
\put(606,525){\rule[-0.175pt]{0.699pt}{0.350pt}}
\put(609,524){\rule[-0.175pt]{0.699pt}{0.350pt}}
\put(612,523){\rule[-0.175pt]{0.699pt}{0.350pt}}
\put(615,522){\rule[-0.175pt]{0.699pt}{0.350pt}}
\put(618,521){\rule[-0.175pt]{0.699pt}{0.350pt}}
\put(621,520){\rule[-0.175pt]{0.657pt}{0.350pt}}
\put(623,519){\rule[-0.175pt]{0.657pt}{0.350pt}}
\put(626,518){\rule[-0.175pt]{0.657pt}{0.350pt}}
\put(629,517){\rule[-0.175pt]{0.657pt}{0.350pt}}
\put(631,516){\rule[-0.175pt]{0.657pt}{0.350pt}}
\put(634,515){\rule[-0.175pt]{0.657pt}{0.350pt}}
\put(637,514){\rule[-0.175pt]{0.657pt}{0.350pt}}
\put(640,513){\rule[-0.175pt]{0.657pt}{0.350pt}}
\put(642,512){\rule[-0.175pt]{0.657pt}{0.350pt}}
\put(645,511){\rule[-0.175pt]{0.657pt}{0.350pt}}
\put(648,510){\rule[-0.175pt]{0.657pt}{0.350pt}}
\put(651,509){\rule[-0.175pt]{0.776pt}{0.350pt}}
\put(654,508){\rule[-0.175pt]{0.776pt}{0.350pt}}
\put(657,507){\rule[-0.175pt]{0.776pt}{0.350pt}}
\put(660,506){\rule[-0.175pt]{0.776pt}{0.350pt}}
\put(663,505){\rule[-0.175pt]{0.776pt}{0.350pt}}
\put(667,504){\rule[-0.175pt]{0.776pt}{0.350pt}}
\put(670,503){\rule[-0.175pt]{0.776pt}{0.350pt}}
\put(673,502){\rule[-0.175pt]{0.776pt}{0.350pt}}
\put(676,501){\rule[-0.175pt]{0.776pt}{0.350pt}}
\put(680,500){\rule[-0.175pt]{0.776pt}{0.350pt}}
\put(683,499){\rule[-0.175pt]{0.776pt}{0.350pt}}
\put(686,498){\rule[-0.175pt]{0.776pt}{0.350pt}}
\put(689,497){\rule[-0.175pt]{0.776pt}{0.350pt}}
\put(692,496){\rule[-0.175pt]{0.776pt}{0.350pt}}
\put(696,495){\rule[-0.175pt]{0.776pt}{0.350pt}}
\put(699,494){\rule[-0.175pt]{0.776pt}{0.350pt}}
\put(702,493){\rule[-0.175pt]{0.776pt}{0.350pt}}
\put(705,492){\rule[-0.175pt]{0.776pt}{0.350pt}}
\put(709,491){\rule[-0.175pt]{0.903pt}{0.350pt}}
\put(712,490){\rule[-0.175pt]{0.903pt}{0.350pt}}
\put(716,489){\rule[-0.175pt]{0.903pt}{0.350pt}}
\put(720,488){\rule[-0.175pt]{0.903pt}{0.350pt}}
\put(724,487){\rule[-0.175pt]{0.903pt}{0.350pt}}
\put(727,486){\rule[-0.175pt]{0.903pt}{0.350pt}}
\put(731,485){\rule[-0.175pt]{0.903pt}{0.350pt}}
\put(735,484){\rule[-0.175pt]{0.903pt}{0.350pt}}
\put(739,483){\rule[-0.175pt]{0.873pt}{0.350pt}}
\put(742,482){\rule[-0.175pt]{0.873pt}{0.350pt}}
\put(746,481){\rule[-0.175pt]{0.873pt}{0.350pt}}
\put(749,480){\rule[-0.175pt]{0.873pt}{0.350pt}}
\put(753,479){\rule[-0.175pt]{0.873pt}{0.350pt}}
\put(757,478){\rule[-0.175pt]{0.873pt}{0.350pt}}
\put(760,477){\rule[-0.175pt]{0.873pt}{0.350pt}}
\put(764,476){\rule[-0.175pt]{0.873pt}{0.350pt}}
\put(768,475){\rule[-0.175pt]{0.998pt}{0.350pt}}
\put(772,474){\rule[-0.175pt]{0.998pt}{0.350pt}}
\put(776,473){\rule[-0.175pt]{0.998pt}{0.350pt}}
\put(780,472){\rule[-0.175pt]{0.998pt}{0.350pt}}
\put(784,471){\rule[-0.175pt]{0.998pt}{0.350pt}}
\put(788,470){\rule[-0.175pt]{0.998pt}{0.350pt}}
\put(792,469){\rule[-0.175pt]{0.998pt}{0.350pt}}
\put(797,468){\rule[-0.175pt]{1.204pt}{0.350pt}}
\put(802,467){\rule[-0.175pt]{1.204pt}{0.350pt}}
\put(807,466){\rule[-0.175pt]{1.204pt}{0.350pt}}
\put(812,465){\rule[-0.175pt]{1.204pt}{0.350pt}}
\put(817,464){\rule[-0.175pt]{1.204pt}{0.350pt}}
\put(822,463){\rule[-0.175pt]{1.204pt}{0.350pt}}
\put(827,462){\rule[-0.175pt]{0.998pt}{0.350pt}}
\put(831,461){\rule[-0.175pt]{0.998pt}{0.350pt}}
\put(835,460){\rule[-0.175pt]{0.998pt}{0.350pt}}
\put(839,459){\rule[-0.175pt]{0.998pt}{0.350pt}}
\put(843,458){\rule[-0.175pt]{0.998pt}{0.350pt}}
\put(847,457){\rule[-0.175pt]{0.998pt}{0.350pt}}
\put(851,456){\rule[-0.175pt]{0.998pt}{0.350pt}}
\put(856,455){\rule[-0.175pt]{1.397pt}{0.350pt}}
\put(861,454){\rule[-0.175pt]{1.397pt}{0.350pt}}
\put(867,453){\rule[-0.175pt]{1.397pt}{0.350pt}}
\put(873,452){\rule[-0.175pt]{1.397pt}{0.350pt}}
\put(879,451){\rule[-0.175pt]{1.397pt}{0.350pt}}
\put(884,450){\rule[-0.175pt]{1.164pt}{0.350pt}}
\put(889,449){\rule[-0.175pt]{1.164pt}{0.350pt}}
\put(894,448){\rule[-0.175pt]{1.164pt}{0.350pt}}
\put(899,447){\rule[-0.175pt]{1.164pt}{0.350pt}}
\put(904,446){\rule[-0.175pt]{1.164pt}{0.350pt}}
\put(909,445){\rule[-0.175pt]{1.164pt}{0.350pt}}
\put(913,444){\rule[-0.175pt]{1.445pt}{0.350pt}}
\put(920,443){\rule[-0.175pt]{1.445pt}{0.350pt}}
\put(926,442){\rule[-0.175pt]{1.445pt}{0.350pt}}
\put(932,441){\rule[-0.175pt]{1.445pt}{0.350pt}}
\put(938,440){\rule[-0.175pt]{1.445pt}{0.350pt}}
\put(944,439){\rule[-0.175pt]{1.397pt}{0.350pt}}
\put(949,438){\rule[-0.175pt]{1.397pt}{0.350pt}}
\put(955,437){\rule[-0.175pt]{1.397pt}{0.350pt}}
\put(961,436){\rule[-0.175pt]{1.397pt}{0.350pt}}
\put(967,435){\rule[-0.175pt]{1.397pt}{0.350pt}}
\put(972,434){\rule[-0.175pt]{1.747pt}{0.350pt}}
\put(980,433){\rule[-0.175pt]{1.747pt}{0.350pt}}
\put(987,432){\rule[-0.175pt]{1.747pt}{0.350pt}}
\put(994,431){\rule[-0.175pt]{1.747pt}{0.350pt}}
\put(1002,430){\rule[-0.175pt]{1.445pt}{0.350pt}}
\put(1008,429){\rule[-0.175pt]{1.445pt}{0.350pt}}
\put(1014,428){\rule[-0.175pt]{1.445pt}{0.350pt}}
\put(1020,427){\rule[-0.175pt]{1.445pt}{0.350pt}}
\put(1026,426){\rule[-0.175pt]{1.445pt}{0.350pt}}
\put(1032,425){\rule[-0.175pt]{1.747pt}{0.350pt}}
\put(1039,424){\rule[-0.175pt]{1.747pt}{0.350pt}}
\put(1046,423){\rule[-0.175pt]{1.747pt}{0.350pt}}
\put(1053,422){\rule[-0.175pt]{1.747pt}{0.350pt}}
\put(1061,421){\rule[-0.175pt]{2.329pt}{0.350pt}}
\put(1070,420){\rule[-0.175pt]{2.329pt}{0.350pt}}
\put(1080,419){\rule[-0.175pt]{2.329pt}{0.350pt}}
\put(1089,418){\rule[-0.175pt]{1.807pt}{0.350pt}}
\put(1097,417){\rule[-0.175pt]{1.807pt}{0.350pt}}
\put(1105,416){\rule[-0.175pt]{1.807pt}{0.350pt}}
\put(1112,415){\rule[-0.175pt]{1.807pt}{0.350pt}}
\put(1120,414){\rule[-0.175pt]{2.329pt}{0.350pt}}
\put(1129,413){\rule[-0.175pt]{2.329pt}{0.350pt}}
\put(1139,412){\rule[-0.175pt]{2.329pt}{0.350pt}}
\put(1148,411){\rule[-0.175pt]{2.329pt}{0.350pt}}
\put(1158,410){\rule[-0.175pt]{2.329pt}{0.350pt}}
\put(1168,409){\rule[-0.175pt]{2.329pt}{0.350pt}}
\put(1177,408){\rule[-0.175pt]{2.329pt}{0.350pt}}
\put(1187,407){\rule[-0.175pt]{2.329pt}{0.350pt}}
\put(1197,406){\rule[-0.175pt]{2.329pt}{0.350pt}}
\put(1206,405){\rule[-0.175pt]{2.409pt}{0.350pt}}
\put(1217,404){\rule[-0.175pt]{2.409pt}{0.350pt}}
\put(1227,403){\rule[-0.175pt]{2.409pt}{0.350pt}}
\put(1237,402){\rule[-0.175pt]{3.493pt}{0.350pt}}
\put(1251,401){\rule[-0.175pt]{3.493pt}{0.350pt}}
\put(1266,400){\rule[-0.175pt]{3.493pt}{0.350pt}}
\put(1280,399){\rule[-0.175pt]{3.493pt}{0.350pt}}
\put(1295,398){\rule[-0.175pt]{2.409pt}{0.350pt}}
\put(1305,397){\rule[-0.175pt]{2.409pt}{0.350pt}}
\put(1315,396){\rule[-0.175pt]{2.409pt}{0.350pt}}
\put(1325,395){\rule[-0.175pt]{3.493pt}{0.350pt}}
\put(1339,394){\rule[-0.175pt]{3.493pt}{0.350pt}}
\end{picture}
\vspace{-.6cm}
\caption{\small\sf Two-loop heavy top corrections to $\drhoc$ in 
units $10^{-4}$.  The upper  line 
represents the leading $O(G_\mu^2 m_t^4)$ correction 
of Ref. [5].}  
\vspace{-4mm}
\end{figure}
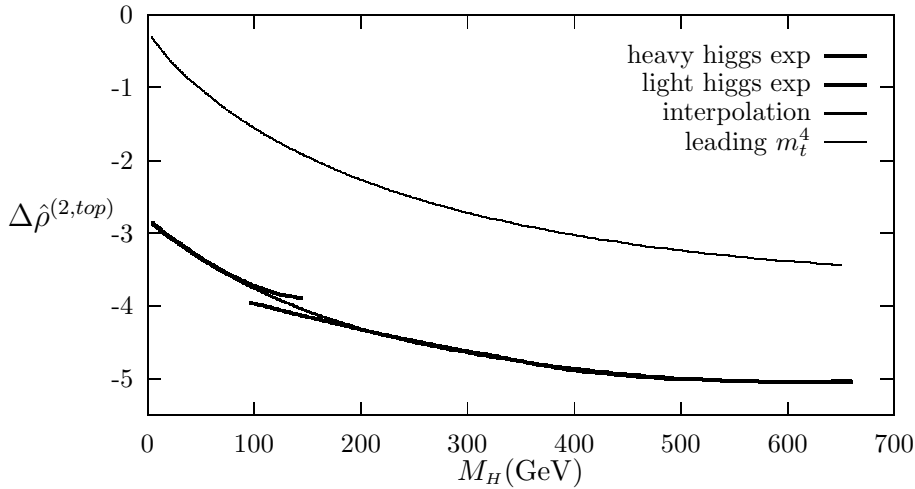
We observe that the correction is extremely small for a small Higgs mass, 
due to large cancellations. We  can naively expect that setting 
the masses of the vector boson different from zero (and so going beyond 
the pure Yukawa theory considered in \cite{barb}) might spoil the 
cancellations and lead to relevant deviations from the upper curve of 
Fig.1 in the light Higgs region.

In addition,   the theoretical \, uncertainty\, coming from unknown
higher order effects is dominated by terms \amtd\ \cite{YB}. 
  Indeed, the  renormalization 
scheme ambiguities and the different resummation options examined in
\cite{YB} led to an estimate of the uncertainty of the theoretical 
predictions which was in a few cases  disturbingly sizable, i.e.
comparable to the present experimental error.
In particular, for $\sineff$, the estimated uncertainty was 
$\delta\,\sineff (th)\lequiv\, 1.4\times 10^{-4}$, 
while the present experimental
average is $\sineff= 0.23165\pm 0.00024$ \cite{war}.
For $\mw$ the uncertainty, $\delta{\mw} (th)\lequiv\,
16$ MeV,  was much  smaller than the  error 
on the present world average, 125 MeV.
 A different analysis based on the explicit two-loop calculation 
of the $\rho$ parameter in low-energy processes has also reached 
very similar conclusions \cite{us}.

Motivated  by the previous observations, the complete 
analytic calculation of the 
two-loop quadratic top effects has been performed  
for the relation between the 
vector boson masses and the muon decay constant $G_\mu$ \cite{physlett}
and for the effective leptonic mixing angle, $\sineff$ \cite{zako,prep}.
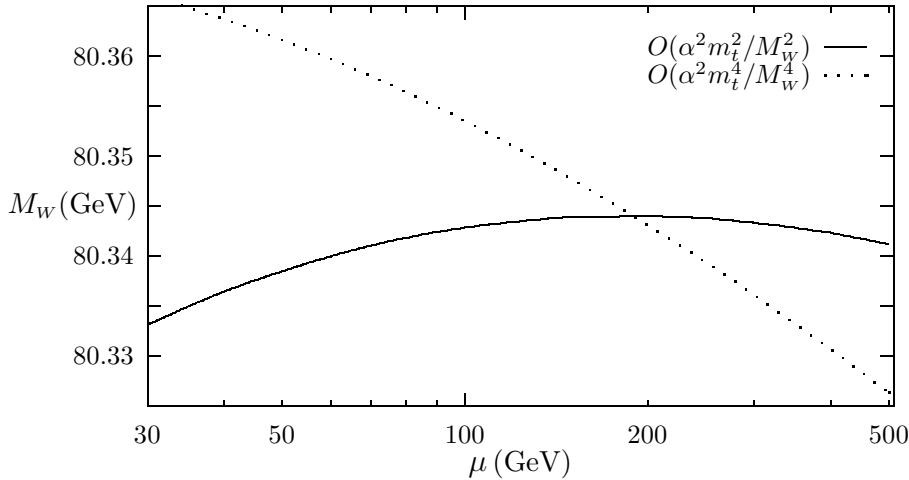
\begin{figure}[t]
\setlength{\unitlength}{0.240900pt}
\ifx\plotpoint\undefined\newsavebox{\plotpoint}\fi
\sbox{\plotpoint}{\rule[-0.175pt]{0.350pt}{0.350pt}}%
\begin{picture}(1500,800)(0,0)
\sbox{\plotpoint}{\rule[-0.175pt]{0.350pt}{0.350pt}}%
\put(264,158){\rule[-0.175pt]{4.818pt}{0.350pt}}
\put(242,158){\makebox(0,0)[r]{}}
\put(1416,158){\rule[-0.175pt]{4.818pt}{0.350pt}}
\put(264,237){\rule[-0.175pt]{4.818pt}{0.350pt}}
\put(242,237){\makebox(0,0)[r]{\ft 80.33}}
\put(1416,237){\rule[-0.175pt]{4.818pt}{0.350pt}}
\put(264,315){\rule[-0.175pt]{4.818pt}{0.350pt}}
\put(242,315){\makebox(0,0)[r]{}}
\put(1416,315){\rule[-0.175pt]{4.818pt}{0.350pt}}
\put(264,394){\rule[-0.175pt]{4.818pt}{0.350pt}}
\put(242,394){\makebox(0,0)[r]{\ft 80.34}}
\put(1416,394){\rule[-0.175pt]{4.818pt}{0.350pt}}
\put(264,472){\rule[-0.175pt]{4.818pt}{0.350pt}}
\put(242,472){\makebox(0,0)[r]{}}
\put(1416,472){\rule[-0.175pt]{4.818pt}{0.350pt}}
\put(264,551){\rule[-0.175pt]{4.818pt}{0.350pt}}
\put(242,551){\makebox(0,0)[r]{\ft 80.35}}
\put(1416,551){\rule[-0.175pt]{4.818pt}{0.350pt}}
\put(264,630){\rule[-0.175pt]{4.818pt}{0.350pt}}
\put(242,630){\makebox(0,0)[r]{}}
\put(1416,630){\rule[-0.175pt]{4.818pt}{0.350pt}}
\put(264,708){\rule[-0.175pt]{4.818pt}{0.350pt}}
\put(242,708){\makebox(0,0)[r]{\ft 80.36}}
\put(1416,708){\rule[-0.175pt]{4.818pt}{0.350pt}}
\put(264,787){\rule[-0.175pt]{4.818pt}{0.350pt}}
\put(242,787){\makebox(0,0)[r]{}}
\put(1416,787){\rule[-0.175pt]{4.818pt}{0.350pt}}
\put(264,158){\rule[-0.175pt]{0.350pt}{2.409pt}}
\put(264,777){\rule[-0.175pt]{0.350pt}{2.409pt}}
\put(264,113){\makebox(0,0){\ft 30}}
\put(383,158){\rule[-0.175pt]{0.350pt}{2.409pt}}
\put(383,777){\rule[-0.175pt]{0.350pt}{2.409pt}}
\put(475,158){\rule[-0.175pt]{0.350pt}{2.409pt}}
\put(475,777){\rule[-0.175pt]{0.350pt}{2.409pt}}
\put(475,113){\makebox(0,0){\ft 50}}
\put(551,158){\rule[-0.175pt]{0.350pt}{2.409pt}}
\put(551,777){\rule[-0.175pt]{0.350pt}{2.409pt}}
\put(614,158){\rule[-0.175pt]{0.350pt}{2.409pt}}
\put(614,777){\rule[-0.175pt]{0.350pt}{2.409pt}}
\put(670,158){\rule[-0.175pt]{0.350pt}{2.409pt}}
\put(670,777){\rule[-0.175pt]{0.350pt}{2.409pt}}
\put(718,158){\rule[-0.175pt]{0.350pt}{2.409pt}}
\put(718,777){\rule[-0.175pt]{0.350pt}{2.409pt}}
\put(762,158){\rule[-0.175pt]{0.350pt}{4.818pt}}
\put(762,113){\makebox(0,0){\ft 100}}
\put(762,767){\rule[-0.175pt]{0.350pt}{4.818pt}}
\put(1049,158){\rule[-0.175pt]{0.350pt}{2.409pt}}
\put(1049,777){\rule[-0.175pt]{0.350pt}{2.409pt}}
\put(1049,113){\makebox(0,0){\ft 200}}
\put(1216,158){\rule[-0.175pt]{0.350pt}{2.409pt}}
\put(1216,777){\rule[-0.175pt]{0.350pt}{2.409pt}}
\put(1336,158){\rule[-0.175pt]{0.350pt}{2.409pt}}
\put(1336,777){\rule[-0.175pt]{0.350pt}{2.409pt}}
\put(1428,158){\rule[-0.175pt]{0.350pt}{2.409pt}}
\put(1428,777){\rule[-0.175pt]{0.350pt}{2.409pt}}
\put(1428,113){\makebox(0,0){\ft 500}}
\put(264,158){\rule[-0.175pt]{282.335pt}{0.350pt}}
\put(1436,158){\rule[-0.175pt]{0.350pt}{151.526pt}}
\put(264,787){\rule[-0.175pt]{282.335pt}{0.350pt}}
\put(45,472){\makebox(0,0)[l]{\shortstack{\small$\mw${(GeV)}}}}
\put(850,68){\makebox(0,0){$\mu$\,{\small (GeV)}}}
\put(264,158){\rule[-0.175pt]{0.350pt}{151.526pt}}
\put(1306,722){\makebox(0,0)[r]{\ft\amtd}}
\put(1328,722){\rule[-0.175pt]{15.899pt}{0.350pt}}
\put(264,286){\usebox{\plotpoint}}
\put(264,286){\rule[-0.175pt]{0.532pt}{0.350pt}}
\put(266,287){\rule[-0.175pt]{0.532pt}{0.350pt}}
\put(268,288){\rule[-0.175pt]{0.532pt}{0.350pt}}
\put(270,289){\rule[-0.175pt]{0.532pt}{0.350pt}}
\put(272,290){\rule[-0.175pt]{0.532pt}{0.350pt}}
\put(275,291){\rule[-0.175pt]{0.532pt}{0.350pt}}
\put(277,292){\rule[-0.175pt]{0.532pt}{0.350pt}}
\put(279,293){\rule[-0.175pt]{0.532pt}{0.350pt}}
\put(281,294){\rule[-0.175pt]{0.532pt}{0.350pt}}
\put(283,295){\rule[-0.175pt]{0.532pt}{0.350pt}}
\put(286,296){\rule[-0.175pt]{0.532pt}{0.350pt}}
\put(288,297){\rule[-0.175pt]{0.532pt}{0.350pt}}
\put(290,298){\rule[-0.175pt]{0.532pt}{0.350pt}}
\put(292,299){\rule[-0.175pt]{0.532pt}{0.350pt}}
\put(294,300){\rule[-0.175pt]{0.532pt}{0.350pt}}
\put(297,301){\rule[-0.175pt]{0.532pt}{0.350pt}}
\put(299,302){\rule[-0.175pt]{0.532pt}{0.350pt}}
\put(301,303){\rule[-0.175pt]{0.532pt}{0.350pt}}
\put(303,304){\rule[-0.175pt]{0.532pt}{0.350pt}}
\put(305,305){\rule[-0.175pt]{0.532pt}{0.350pt}}
\put(308,306){\rule[-0.175pt]{0.532pt}{0.350pt}}
\put(310,307){\rule[-0.175pt]{0.532pt}{0.350pt}}
\put(312,308){\rule[-0.175pt]{0.532pt}{0.350pt}}
\put(314,309){\rule[-0.175pt]{0.532pt}{0.350pt}}
\put(316,310){\rule[-0.175pt]{0.532pt}{0.350pt}}
\put(319,311){\rule[-0.175pt]{0.532pt}{0.350pt}}
\put(321,312){\rule[-0.175pt]{0.532pt}{0.350pt}}
\put(323,313){\rule[-0.175pt]{0.532pt}{0.350pt}}
\put(325,314){\rule[-0.175pt]{0.532pt}{0.350pt}}
\put(328,315){\rule[-0.175pt]{0.576pt}{0.350pt}}
\put(330,316){\rule[-0.175pt]{0.576pt}{0.350pt}}
\put(332,317){\rule[-0.175pt]{0.576pt}{0.350pt}}
\put(335,318){\rule[-0.175pt]{0.576pt}{0.350pt}}
\put(337,319){\rule[-0.175pt]{0.576pt}{0.350pt}}
\put(339,320){\rule[-0.175pt]{0.576pt}{0.350pt}}
\put(342,321){\rule[-0.175pt]{0.576pt}{0.350pt}}
\put(344,322){\rule[-0.175pt]{0.576pt}{0.350pt}}
\put(347,323){\rule[-0.175pt]{0.576pt}{0.350pt}}
\put(349,324){\rule[-0.175pt]{0.576pt}{0.350pt}}
\put(351,325){\rule[-0.175pt]{0.576pt}{0.350pt}}
\put(354,326){\rule[-0.175pt]{0.576pt}{0.350pt}}
\put(356,327){\rule[-0.175pt]{0.576pt}{0.350pt}}
\put(359,328){\rule[-0.175pt]{0.576pt}{0.350pt}}
\put(361,329){\rule[-0.175pt]{0.576pt}{0.350pt}}
\put(363,330){\rule[-0.175pt]{0.576pt}{0.350pt}}
\put(366,331){\rule[-0.175pt]{0.576pt}{0.350pt}}
\put(368,332){\rule[-0.175pt]{0.576pt}{0.350pt}}
\put(371,333){\rule[-0.175pt]{0.576pt}{0.350pt}}
\put(373,334){\rule[-0.175pt]{0.576pt}{0.350pt}}
\put(375,335){\rule[-0.175pt]{0.576pt}{0.350pt}}
\put(378,336){\rule[-0.175pt]{0.576pt}{0.350pt}}
\put(380,337){\rule[-0.175pt]{0.576pt}{0.350pt}}
\put(382,338){\rule[-0.175pt]{0.656pt}{0.350pt}}
\put(385,339){\rule[-0.175pt]{0.656pt}{0.350pt}}
\put(388,340){\rule[-0.175pt]{0.656pt}{0.350pt}}
\put(391,341){\rule[-0.175pt]{0.656pt}{0.350pt}}
\put(393,342){\rule[-0.175pt]{0.656pt}{0.350pt}}
\put(396,343){\rule[-0.175pt]{0.656pt}{0.350pt}}
\put(399,344){\rule[-0.175pt]{0.656pt}{0.350pt}}
\put(402,345){\rule[-0.175pt]{0.656pt}{0.350pt}}
\put(404,346){\rule[-0.175pt]{0.656pt}{0.350pt}}
\put(407,347){\rule[-0.175pt]{0.656pt}{0.350pt}}
\put(410,348){\rule[-0.175pt]{0.656pt}{0.350pt}}
\put(412,349){\rule[-0.175pt]{0.656pt}{0.350pt}}
\put(415,350){\rule[-0.175pt]{0.656pt}{0.350pt}}
\put(418,351){\rule[-0.175pt]{0.656pt}{0.350pt}}
\put(421,352){\rule[-0.175pt]{0.656pt}{0.350pt}}
\put(423,353){\rule[-0.175pt]{0.656pt}{0.350pt}}
\put(426,354){\rule[-0.175pt]{0.656pt}{0.350pt}}
\put(429,355){\rule[-0.175pt]{0.656pt}{0.350pt}}
\put(432,356){\rule[-0.175pt]{0.741pt}{0.350pt}}
\put(435,357){\rule[-0.175pt]{0.741pt}{0.350pt}}
\put(438,358){\rule[-0.175pt]{0.741pt}{0.350pt}}
\put(441,359){\rule[-0.175pt]{0.741pt}{0.350pt}}
\put(444,360){\rule[-0.175pt]{0.741pt}{0.350pt}}
\put(447,361){\rule[-0.175pt]{0.741pt}{0.350pt}}
\put(450,362){\rule[-0.175pt]{0.741pt}{0.350pt}}
\put(453,363){\rule[-0.175pt]{0.741pt}{0.350pt}}
\put(456,364){\rule[-0.175pt]{0.741pt}{0.350pt}}
\put(459,365){\rule[-0.175pt]{0.741pt}{0.350pt}}
\put(462,366){\rule[-0.175pt]{0.741pt}{0.350pt}}
\put(465,367){\rule[-0.175pt]{0.741pt}{0.350pt}}
\put(468,368){\rule[-0.175pt]{0.741pt}{0.350pt}}
\put(471,369){\rule[-0.175pt]{0.741pt}{0.350pt}}
\put(475,370){\rule[-0.175pt]{0.741pt}{0.350pt}}
\put(478,371){\rule[-0.175pt]{0.741pt}{0.350pt}}
\put(481,372){\rule[-0.175pt]{0.741pt}{0.350pt}}
\put(484,373){\rule[-0.175pt]{0.741pt}{0.350pt}}
\put(487,374){\rule[-0.175pt]{0.741pt}{0.350pt}}
\put(490,375){\rule[-0.175pt]{0.741pt}{0.350pt}}
\put(493,376){\rule[-0.175pt]{0.741pt}{0.350pt}}
\put(496,377){\rule[-0.175pt]{0.741pt}{0.350pt}}
\put(499,378){\rule[-0.175pt]{0.741pt}{0.350pt}}
\put(502,379){\rule[-0.175pt]{0.741pt}{0.350pt}}
\put(505,380){\rule[-0.175pt]{0.741pt}{0.350pt}}
\put(508,381){\rule[-0.175pt]{0.741pt}{0.350pt}}
\put(511,382){\rule[-0.175pt]{0.741pt}{0.350pt}}
\put(514,383){\rule[-0.175pt]{0.831pt}{0.350pt}}
\put(518,384){\rule[-0.175pt]{0.831pt}{0.350pt}}
\put(521,385){\rule[-0.175pt]{0.831pt}{0.350pt}}
\put(525,386){\rule[-0.175pt]{0.831pt}{0.350pt}}
\put(528,387){\rule[-0.175pt]{0.831pt}{0.350pt}}
\put(532,388){\rule[-0.175pt]{0.831pt}{0.350pt}}
\put(535,389){\rule[-0.175pt]{0.831pt}{0.350pt}}
\put(539,390){\rule[-0.175pt]{0.831pt}{0.350pt}}
\put(542,391){\rule[-0.175pt]{0.831pt}{0.350pt}}
\put(546,392){\rule[-0.175pt]{0.831pt}{0.350pt}}
\put(549,393){\rule[-0.175pt]{0.831pt}{0.350pt}}
\put(552,394){\rule[-0.175pt]{0.831pt}{0.350pt}}
\put(556,395){\rule[-0.175pt]{0.831pt}{0.350pt}}
\put(559,396){\rule[-0.175pt]{0.831pt}{0.350pt}}
\put(563,397){\rule[-0.175pt]{0.831pt}{0.350pt}}
\put(566,398){\rule[-0.175pt]{0.831pt}{0.350pt}}
\put(570,399){\rule[-0.175pt]{0.831pt}{0.350pt}}
\put(573,400){\rule[-0.175pt]{0.831pt}{0.350pt}}
\put(577,401){\rule[-0.175pt]{0.831pt}{0.350pt}}
\put(580,402){\rule[-0.175pt]{0.831pt}{0.350pt}}
\put(584,403){\rule[-0.175pt]{0.964pt}{0.350pt}}
\put(588,404){\rule[-0.175pt]{0.964pt}{0.350pt}}
\put(592,405){\rule[-0.175pt]{0.964pt}{0.350pt}}
\put(596,406){\rule[-0.175pt]{0.964pt}{0.350pt}}
\put(600,407){\rule[-0.175pt]{0.964pt}{0.350pt}}
\put(604,408){\rule[-0.175pt]{0.964pt}{0.350pt}}
\put(608,409){\rule[-0.175pt]{0.964pt}{0.350pt}}
\put(612,410){\rule[-0.175pt]{0.964pt}{0.350pt}}
\put(616,411){\rule[-0.175pt]{0.964pt}{0.350pt}}
\put(620,412){\rule[-0.175pt]{0.964pt}{0.350pt}}
\put(624,413){\rule[-0.175pt]{0.964pt}{0.350pt}}
\put(628,414){\rule[-0.175pt]{0.964pt}{0.350pt}}
\put(632,415){\rule[-0.175pt]{1.144pt}{0.350pt}}
\put(636,416){\rule[-0.175pt]{1.144pt}{0.350pt}}
\put(641,417){\rule[-0.175pt]{1.144pt}{0.350pt}}
\put(646,418){\rule[-0.175pt]{1.144pt}{0.350pt}}
\put(651,419){\rule[-0.175pt]{1.144pt}{0.350pt}}
\put(655,420){\rule[-0.175pt]{1.144pt}{0.350pt}}
\put(660,421){\rule[-0.175pt]{1.144pt}{0.350pt}}
\put(665,422){\rule[-0.175pt]{1.144pt}{0.350pt}}
\put(670,423){\rule[-0.175pt]{1.204pt}{0.350pt}}
\put(675,424){\rule[-0.175pt]{1.204pt}{0.350pt}}
\put(680,425){\rule[-0.175pt]{1.204pt}{0.350pt}}
\put(685,426){\rule[-0.175pt]{1.204pt}{0.350pt}}
\put(690,427){\rule[-0.175pt]{1.204pt}{0.350pt}}
\put(695,428){\rule[-0.175pt]{1.204pt}{0.350pt}}
\put(700,429){\rule[-0.175pt]{1.445pt}{0.350pt}}
\put(706,430){\rule[-0.175pt]{1.445pt}{0.350pt}}
\put(712,431){\rule[-0.175pt]{1.445pt}{0.350pt}}
\put(718,432){\rule[-0.175pt]{1.445pt}{0.350pt}}
\put(724,433){\rule[-0.175pt]{1.365pt}{0.350pt}}
\put(729,434){\rule[-0.175pt]{1.365pt}{0.350pt}}
\put(735,435){\rule[-0.175pt]{1.365pt}{0.350pt}}
\put(741,436){\rule[-0.175pt]{1.686pt}{0.350pt}}
\put(748,437){\rule[-0.175pt]{1.686pt}{0.350pt}}
\put(755,438){\rule[-0.175pt]{1.686pt}{0.350pt}}
\put(762,439){\rule[-0.175pt]{1.879pt}{0.350pt}}
\put(769,440){\rule[-0.175pt]{1.879pt}{0.350pt}}
\put(777,441){\rule[-0.175pt]{1.879pt}{0.350pt}}
\put(785,442){\rule[-0.175pt]{1.879pt}{0.350pt}}
\put(793,443){\rule[-0.175pt]{1.879pt}{0.350pt}}
\put(800,444){\rule[-0.175pt]{2.168pt}{0.350pt}}
\put(810,445){\rule[-0.175pt]{2.168pt}{0.350pt}}
\put(819,446){\rule[-0.175pt]{2.168pt}{0.350pt}}
\put(828,447){\rule[-0.175pt]{2.168pt}{0.350pt}}
\put(837,448){\rule[-0.175pt]{2.730pt}{0.350pt}}
\put(848,449){\rule[-0.175pt]{2.730pt}{0.350pt}}
\put(859,450){\rule[-0.175pt]{2.730pt}{0.350pt}}
\put(870,451){\rule[-0.175pt]{2.710pt}{0.350pt}}
\put(882,452){\rule[-0.175pt]{2.710pt}{0.350pt}}
\put(893,453){\rule[-0.175pt]{2.710pt}{0.350pt}}
\put(904,454){\rule[-0.175pt]{2.710pt}{0.350pt}}
\put(916,455){\rule[-0.175pt]{9.636pt}{0.350pt}}
\put(956,456){\rule[-0.175pt]{11.804pt}{0.350pt}}
\put(1005,457){\rule[-0.175pt]{16.381pt}{0.350pt}}
\put(1073,456){\rule[-0.175pt]{5.782pt}{0.350pt}}
\put(1097,455){\rule[-0.175pt]{5.300pt}{0.350pt}}
\put(1119,454){\rule[-0.175pt]{5.300pt}{0.350pt}}
\put(1141,453){\rule[-0.175pt]{2.581pt}{0.350pt}}
\put(1151,452){\rule[-0.175pt]{2.581pt}{0.350pt}}
\put(1162,451){\rule[-0.175pt]{2.581pt}{0.350pt}}
\put(1173,450){\rule[-0.175pt]{2.581pt}{0.350pt}}
\put(1183,449){\rule[-0.175pt]{2.581pt}{0.350pt}}
\put(1194,448){\rule[-0.175pt]{2.581pt}{0.350pt}}
\put(1205,447){\rule[-0.175pt]{2.581pt}{0.350pt}}
\put(1215,446){\rule[-0.175pt]{1.927pt}{0.350pt}}
\put(1224,445){\rule[-0.175pt]{1.927pt}{0.350pt}}
\put(1232,444){\rule[-0.175pt]{1.927pt}{0.350pt}}
\put(1240,443){\rule[-0.175pt]{1.927pt}{0.350pt}}
\put(1248,442){\rule[-0.175pt]{1.927pt}{0.350pt}}
\put(1256,441){\rule[-0.175pt]{1.927pt}{0.350pt}}
\put(1264,440){\rule[-0.175pt]{1.927pt}{0.350pt}}
\put(1272,439){\rule[-0.175pt]{1.927pt}{0.350pt}}
\put(1280,438){\rule[-0.175pt]{1.686pt}{0.350pt}}
\put(1287,437){\rule[-0.175pt]{1.686pt}{0.350pt}}
\put(1294,436){\rule[-0.175pt]{1.686pt}{0.350pt}}
\put(1301,435){\rule[-0.175pt]{1.686pt}{0.350pt}}
\put(1308,434){\rule[-0.175pt]{1.686pt}{0.350pt}}
\put(1315,433){\rule[-0.175pt]{1.686pt}{0.350pt}}
\put(1322,432){\rule[-0.175pt]{1.686pt}{0.350pt}}
\put(1329,431){\rule[-0.175pt]{1.686pt}{0.350pt}}
\put(1336,430){\rule[-0.175pt]{1.285pt}{0.350pt}}
\put(1341,429){\rule[-0.175pt]{1.285pt}{0.350pt}}
\put(1346,428){\rule[-0.175pt]{1.285pt}{0.350pt}}
\put(1352,427){\rule[-0.175pt]{1.285pt}{0.350pt}}
\put(1357,426){\rule[-0.175pt]{1.285pt}{0.350pt}}
\put(1362,425){\rule[-0.175pt]{1.285pt}{0.350pt}}
\put(1368,424){\rule[-0.175pt]{1.285pt}{0.350pt}}
\put(1373,423){\rule[-0.175pt]{1.285pt}{0.350pt}}
\put(1378,422){\rule[-0.175pt]{1.285pt}{0.350pt}}
\put(1384,421){\rule[-0.175pt]{1.178pt}{0.350pt}}
\put(1388,420){\rule[-0.175pt]{1.178pt}{0.350pt}}
\put(1393,419){\rule[-0.175pt]{1.178pt}{0.350pt}}
\put(1398,418){\rule[-0.175pt]{1.178pt}{0.350pt}}
\put(1403,417){\rule[-0.175pt]{1.178pt}{0.350pt}}
\put(1408,416){\rule[-0.175pt]{1.178pt}{0.350pt}}
\put(1413,415){\rule[-0.175pt]{1.178pt}{0.350pt}}
\put(1418,414){\rule[-0.175pt]{1.178pt}{0.350pt}}
\put(1423,413){\rule[-0.175pt]{1.178pt}{0.350pt}}
\sbox{\plotpoint}{\rule[-0.250pt]{0.500pt}{0.500pt}}%
\put(1306,677){\makebox(0,0)[r]{\ft\amtq}}
\put(1328,677){\usebox{\plotpoint}}
\put(1348,677){\usebox{\plotpoint}}
\put(1369,677){\usebox{\plotpoint}}
\put(1390,677){\usebox{\plotpoint}}
\put(1394,677){\usebox{\plotpoint}}
\put(317,787){\usebox{\plotpoint}}
\put(336,781){\usebox{\plotpoint}}
\put(356,774){\usebox{\plotpoint}}
\put(376,768){\usebox{\plotpoint}}
\put(396,761){\usebox{\plotpoint}}
\put(415,755){\usebox{\plotpoint}}
\put(435,748){\usebox{\plotpoint}}
\put(454,741){\usebox{\plotpoint}}
\put(474,733){\usebox{\plotpoint}}
\put(493,726){\usebox{\plotpoint}}
\put(512,718){\usebox{\plotpoint}}
\put(532,711){\usebox{\plotpoint}}
\put(551,703){\usebox{\plotpoint}}
\put(570,695){\usebox{\plotpoint}}
\put(589,687){\usebox{\plotpoint}}
\put(608,679){\usebox{\plotpoint}}
\put(627,670){\usebox{\plotpoint}}
\put(646,662){\usebox{\plotpoint}}
\put(665,653){\usebox{\plotpoint}}
\put(684,644){\usebox{\plotpoint}}
\put(702,635){\usebox{\plotpoint}}
\put(721,626){\usebox{\plotpoint}}
\put(740,617){\usebox{\plotpoint}}
\put(758,607){\usebox{\plotpoint}}
\put(777,598){\usebox{\plotpoint}}
\put(795,588){\usebox{\plotpoint}}
\put(814,579){\usebox{\plotpoint}}
\put(832,569){\usebox{\plotpoint}}
\put(850,559){\usebox{\plotpoint}}
\put(868,549){\usebox{\plotpoint}}
\put(886,539){\usebox{\plotpoint}}
\put(904,529){\usebox{\plotpoint}}
\put(922,518){\usebox{\plotpoint}}
\put(940,508){\usebox{\plotpoint}}
\put(958,497){\usebox{\plotpoint}}
\put(976,486){\usebox{\plotpoint}}
\put(994,476){\usebox{\plotpoint}}
\put(1012,465){\usebox{\plotpoint}}
\put(1029,454){\usebox{\plotpoint}}
\put(1047,443){\usebox{\plotpoint}}
\put(1064,432){\usebox{\plotpoint}}
\put(1082,421){\usebox{\plotpoint}}
\put(1099,410){\usebox{\plotpoint}}
\put(1117,399){\usebox{\plotpoint}}
\put(1134,387){\usebox{\plotpoint}}
\put(1152,376){\usebox{\plotpoint}}
\put(1169,364){\usebox{\plotpoint}}
\put(1186,353){\usebox{\plotpoint}}
\put(1203,341){\usebox{\plotpoint}}
\put(1220,329){\usebox{\plotpoint}}
\put(1237,317){\usebox{\plotpoint}}
\put(1254,305){\usebox{\plotpoint}}
\put(1271,293){\usebox{\plotpoint}}
\put(1288,281){\usebox{\plotpoint}}
\put(1305,269){\usebox{\plotpoint}}
\put(1322,257){\usebox{\plotpoint}}
\put(1339,245){\usebox{\plotpoint}}
\put(1355,233){\usebox{\plotpoint}}
\put(1372,220){\usebox{\plotpoint}}
\put(1388,208){\usebox{\plotpoint}}
\put(1405,195){\usebox{\plotpoint}}
\put(1422,183){\usebox{\plotpoint}}
\put(1428,179){\usebox{\plotpoint}}
\end{picture}
\vspace{-11mm}
\caption{\sf\small Scale dependence of the prediction for $\mw$ 
in the $\msbar$
scheme for $\mt=180$GeV, $\mh=300$GeV, including only the leading
\gmtq\ correction (dotted curve)
or all the available two-loop contributions, through
\amtd\ (solid curve).}
\label{mudep}
\vspace{-3mm}
\end{figure}
 Corrections to the 
 $Z^0$ decay width are also under study.
All these calculations have been done in 
the $\msbar$ scheme introduced in \cite{msbar}, i.e. using $\msbar$ 
couplings and on-shell masses, a  particularly convenient framework. 
Besides the leading \gmtq\ result, I have displayed in Fig.1 the
two-loop heavy top correction to $\hat{\rho}$ up to \amtd. 
In the $\msbar$ scheme, $\hat{\rho}$ is defined as 
$\mw^2/\mz^2 \ 1/\cos^2\hat{\theta}_{\msbar}(\mz)$, 
and represents the most obvious 
process-independent analogue of the low-energy $\rho$ parameter. 
The deviation from the leading \gmtq\ result is striking.

In order to gauge the impact of the new calculation, however, 
it is best to 
consider   physical observables, and to study the scheme and scale 
dependence of their predictions from $\alpha$, $G_\mu$, and $\mz$.
 I will do that for the two most relevant 
precision quantities: the mass of the $W$ boson and $\sineff$.
Concerning the scale dependence of the $\msbar$ predictions, 
the situation is 
exemplified in Fig.\ref{mudep}. Over a wide range of $\mu$ values, the
scale dependence of  $\mw$ 
is  significantly reduced by the inclusion
of the \amtd\ contribution.

\renewcommand{\arraystretch}{1.2}
\begin{table}[t] 
\[
\begin{array}{|c||c|c||c|c|}\hline
\mh & \Delta \,{s^2}_{eff}^{lead}  
&\Delta \,{s^2}_{eff}  & \Delta\mw^{lead} & \Delta\mw  \\  \hline\hline
65  &-0.90 & -0.14 & 8.4 &1.5  \\ \hline
100  & -0.90& -0.12 & 8.2& 1.3\\ \hline
300   & -0.87& -0.08 &7.6 &0.5 \\ \hline
600  & -0.83&-0.05   &7.0 &0.1 \\ \hline
1000 & -0.79&-0.03 & 6.5 &-0.3 \\ \hline
\end{array}            
\]
\caption{\sf\small Scheme 
dependence of the prediction of $\sineff$ before and 
after the inclusion of the new \amtd\ correction for $\mt=175$GeV.
$\Delta s^2_{eff}$ is in units $10^{-4}$,  $\Delta\mw$ in MeV and 
$\mh$ in GeV.}
\vspace{-2mm}
\end{table}
After translating  \cite{prep} the results of the two-loop 
calculation into the on-shell (OS)
scheme \cite{si80},  I have compared the 
predictions for $\mw$ and $\sineff$ 
in the $\msbar$  and  OS scheme,
before and after the inclusion of the \amtd\ contribution.
 The results  for one particular OS implementation
are shown in Table 1,
where $\Delta s_{eff}^2\equiv \sineff(\msbar)-
\sineff(\rm OS)$ and $\Delta\mw\equiv \mw(\msbar) - \mw(\rm OS)$.
 As expected, the scheme dependence
of the predicted values of $\sineff$ and $\mw$ is drastically reduced.
After considering alternative OS options,
we can safely conclude that the inclusion of the \amtd\ correction reduces
the scheme dependence in the two cases considered 
by  at least a factor corresponding to the expansion 
parameter $\mw^2/\mt^2\approx 0.2$.

Finally, let me consider the impact of  the new calculations on 
the precise predictions of $\sineff$ and $\mw$. 
It is clear that  the shifts
induced depend very strongly on the actual implementations
of OS and $\msbar$ scheme. The results shown in Table \ref{impact}
 refer to a typical $\msbar$ implementation \cite{zako,msbar}
and to two
different OS implementations, OSI and OSII.
Using $\delta\mw\equiv
\mw - \mw^{lead}$ and $\delta \,s^2_{eff}\equiv
\delta s^2_{eff} - \delta {s^2}_{eff}^{lead}$ 
for the shifts induced by the \amtd\ corrections on $\mw$ and $\sineff$,
we see that $\delta\mw$ and $\delta\, s^2_{eff}$ 
tend to be  larger in the light Higgs region, 
and that  $\delta\, s^2_{eff}$ 
can be quite sizable, more than 1$\times 10^{-4}$.
\begin{table}[t] 
\[
\begin{array}{|c||c|c|c||c|c|c|}\hline
\mh  & \delta \,{s^2}_{eff}^{\rm OSI}  
& \delta \,{s^2}_{eff}^{\rm OSII}  
&\delta \,{s^2}_{eff}^{\msbar}& \delta\mw^{\rm OSI} & \delta\mw^{\rm OSII} 
&\delta\mw^{\msbar} 
\\  \hline\hline
65  &0.04& 1.56 &0.80 &  -6.5& -14.5 &-13.4 \\ \hline
100 & -0.02& 1.27 &0.76 & -5.9 &-12.7 &-12.9 \\ \hline
300   & -0.14& 0.35 &0.65 & -4.1 &-6.7 &-11.1 \\ \hline
600   & -0.20& -0.33&0.58  & -2.6 &-1.9 &-9.5  \\ \hline
1000 & -0.30 &-0.93& 0.46  & -0.5 &2.9 &-7.3 \\ \hline
\end{array}            
\]
\caption{\sf\small Shifts induced  by the \amtd\ corrections
 in the OS and $\msbar$ scheme for $\mt=175$GeV.
$\delta s^2_{eff}$ is in units $10^{-4}$,  $\delta\mw$ in MeV.}
\label{impact}
\vspace{-3mm}
\end{table}
 Because of the sign of the shifts, in general the \amtd\ correction 
further enhances the screening of the top quark
 contribution by higher order effects.

In summary, two-loop \ew\ $\mt^2$ effects are now available
in analytic form
for the main precision observables in {\em both} $\msbar$ and OS 
frameworks. The new contributions consistently reduce the scheme
and scale dependence of the predictions by {\em at least}
a factor $\mw^2/\mt^2\approx 0.2$, suggesting 
a relevant improvement in the 
theoretical accuracy. The impact on 
the value of the effective sine can be sizable, up to 1.5$\times 10^{-4}$,
but it is highly sensitive to  the scheme adopted.

\vspace{.4cm}
I wish to thank the organizers for the excellent organization and
the pleasant atmosphere during the Symposium.
Useful conversations with G. Degrassi, W. Hollik, and A. Sirlin are 
gratefully acknowledged.

\end{document}